\documentclass[10pt]{article}
\usepackage{geometry}
\usepackage{amsfonts}
\usepackage{amsmath}
\usepackage{amssymb}
\usepackage{cancel}
\usepackage{mathrsfs}
\usepackage{wasysym}
\usepackage{cite}
\usepackage{bbm}
\usepackage{bm}
\usepackage{graphicx}
\usepackage[labelfont=small]{caption}
\usepackage{hyperref}
\usepackage[usenames,dvipsnames]{color}
\usepackage{wrapfig}
\usepackage{comment}
\usepackage{multicol}

\usepackage[LGR,T1]{fontenc}
\usepackage[utf8]{inputenc}   

\hypersetup{
colorlinks=true,         
linkcolor=Red,          
citecolor=blue,           
urlcolor=Cyan          
}
\title{\vspace{-1 cm} \textbf{T-Minkowski noncommutative spacetimes I:\\
Poincar\'e groups, differential calculi and braiding}}
\author{Flavio Mercati\footnote{flavio.mercati@gmail.com}
\vspace{12pt}
\\
Departamento de F\'isica, Universidad de Burgos, 09001 Burgos, Spain.
}

\newcommand{\sfrac}[2]{ {\textstyle \frac{#1}{#2}}}

\newcommand{\1}{\hat{1}}
\newcommand{\Q}{4}

\newcommand{\x}{\hat{x}}
\newcommand{\A}{\hat{a}}
\newcommand{\La}{\hat{\Lambda}}

\usepackage{tocloft}

\begin{document}

\maketitle

\begin{abstract}
This paper introduces and investigates a class of noncommutative spacetimes that I will call ``T-Minkowski,'' whose quantum Poincaré group of isometries exhibits unique and physically motivated characteristics. Notably, the coordinates on the Lorentz subgroup remain commutative, while the deformation is confined to the translations (hence the T in the name), which act like an integrable set of vector fields on the Lorentz group. This is similar to Majid's bicrossproduct construction, although my approach allows the description of spacetimes with commutators that include a constant matrix as well as terms that are linear in the coordinates (the resulting structure is that of a centrally-extended Lie algebra). Moreover, I require that one can define a covariant braided tensor product representation of the quantum Poincar\'e group, describing the algebra of N-points. This also implies that a 4-dimensional bicovariant differential calculus exists on the noncommutative spacetime. The resulting models can all be described in terms of a numerical triangular R-matrix through RTT relations (as well as RXX, RXY and RXdX relations for the homogeneous spacetime, the braiding and the differential calculus). The R-matrices I find are in one-to-one correspondence with the triangular r-matrices on the Poincar\'e group without quadratic terms in the Lorentz generators. These have been classified, up to automorphisms,  by Zakrzewski, and amount to 16 inequivalent models. This paper is the first of a series, focusing on the identification of all the quantum Poincar\'e groups that are allowed by my assumptions, as well as the associated quantum homogeneous spacetimes, differential calculi and braiding constructions.
\end{abstract}

\vspace{2cm}

\tableofcontents

\newpage

\subsection*{Notations used in the paper}

\begin{itemize}
\item The Einstein summation convention is assumed. Greek indices $\alpha, \beta , \dots \mu,\nu,\dots$ will run from $0$ to $3$. Lowercase Latin indices from the middle of the alphabet, $i,j,k,l,\dots$ will run from $1$ to $3$. The ones from the beginning of the alphabet, $a,b,c,\dots$ will be natural numbers. Uppercase Latin indices $A,B,C, \dots, I,J,K,\dots$ will run from $0$ to $4$.

\item Symmetrization and antisymmetrization of indices will be done with a $1/2$ weight:
$$
T^{(\mu\nu)} = {\sfrac 1 2} (T^{\mu\nu} + T^{\nu\mu} ) \,,
\qquad
T^{[\mu\nu]} = {\sfrac 1 2} (T^{\mu\nu} - T^{\nu\mu} ) \,.
$$

\item A hat on a symbol, $\hat f$, indicates that $\hat f$ is an element of a not-necessarily-commutative algebra.

\item $\eta_{\mu\nu}$ will indicate a \textit{numerical,} constant symmetric spacetime metric. The Minkowski metric in Cartesian coordinates, with the mostly-positive signature, $\eta_{\mu\nu} = \text{diag}(-1,1,1,1)$, will be assumed unless specified otherwise.

\end{itemize}

\section{Introduction}

This paper is the first installment of a series dedicated to establishing a robust framework for Quantum Field Theory (QFT) on noncommutative spacetimes. The approach I will outline here is neatly delimited by a set of assumptions, which are well motivated by physical and simplicity arguments. In this first paper, I will identify 16 classes of models that are singled out by my assumptions. These are all 4-dimensional noncommutative spacetimes, each invariant under its own quantum-group deformation of the Poincar\'e group. As far as I am aware of, only two of these 16 cases have been studied in the past in any detail, and this series of paper hopes to initiate a detailed analysis of the physical consequences of all of them.

QFT on noncommutative spacetimes  was proposed almost eighty years ago as a way to regularize the ultraviolet divergences of quantum electrodynamics~\cite{Snyder1947}. The idea that spacetime coordinates might be quantized seems like a natural extension of the quantization program~\cite{GamowBook}, and the emergence of the Planck length, $\ell_P=\sqrt{{G\hbar}/{c^3}} \sim 1.6 \times 10^{-35} \, m$, as the scale of breakdown of quantum General Relativity as an effective field theory suggests that Quantum Gravity imposes limits on the localizability of spacetime events~\cite{Hossenfelder:2012jw}, which may require replacing its standard smooth structure  with something new when probing scales that approach $\ell_P$.
As the tools of classical differential topology and Riemannian geometry are insufficient to describe geometrical objects that lack the notion of infinitesimal points,  the study of noncommutative spaces requires a purely algebraic formulation. Thanks to the work of Connes, Woronowicz and Drinfel'd~\cite{Connes1985,Woronowicz1987_1,Woronowicz1987_2}, we now have algebraic formulations of all the basic tools of geometry and topology. These allow one, in principle, to do physics on such spaces. Various aspects of noncommutative geometry have found applications in several fields of physics beyond the (commutative) Standard Model: for example, 2+1-dimensional Quantum Gravity~\cite{Matschull:1997du,Freidel:2005me}, String Theory~\cite{veneziano1986stringy,witten1986noncommutative,Seiberg:1999vs,deBoer:2003dpn} and attempts at explaining the structure of the Standard Model~\cite{connes1989particle,chamseddine1997spectral}.

The appearance of a ``scale of fuzziness'' of spacetime, $\ell_P$, raises a pressing issue:  the fate of local Lorentz invariance, which might seem incompatible with a fundamental limitation to localizability. For instance, nontrivial commutation relations between some spacetime coordinates cannot be an observer-independent notion under standard Lorentz/Poincar\'e transformations: two different reference frames are bound to disagree on which coordinate does not commute with which, and on the structure of the consequent uncertainty relations. This raises the question of Lorentz Invariance Violations, which are extremely  well constrained experimentally~\cite{Mattingly:2005re,Bolmont:2022yad}, and seem to exclude most types of Lorentz-breaking noncommutativity down to scales well beyond the Planck length.

Noncommutative geometry offers a way out of this impasse, in the form of \textit{quantum symmetry groups} and \textit{quantum homogeneous spaces.} This is a notion that emerged naturally in the study of noncommutative (differential) geometry: some noncommutative spacetimes admit a generalized notion of group of symmetries they are invariant under, in an appropriate sense. These are called \textit{quantum groups}~\cite{majid_1995,majid_2002,Chari}, and are described in terms of Hopf algebras~\cite{Szabo:2001kg,aschieri2011noncommutative,Wulkenhaar2019}. A noncommutative \textit{spacetime} that is also the homogeneous space of a quantum group implements, in a novel way,  the principle of relativity: all inertial frames are equivalent, and the uncertainty relations implied by noncommutativity appear the same in all reference frames. This is achieved at the cost of renouncing the classical notion of transformation between reference frames: the transformation parameters, which were coordinates on a Lie group manifold in the commutative settings, are replaced by elements of a noncommutative algebra, and the points on the group manifold have to be replaced with the states on this algebra, which will have some degree of fuzziness. The idea that the Planck length might enter ultrashort-scale physics as a new invariant scale, which is preserved by altering the transformation rules between reference frames, similarly to how the speed of light remains constant in Special Relativity, has been conjectured more than 20 years ago~\cite{Amelino-Camelia:2000stu}. The apparent tension between the existence of a noncommutativity length scale and Lorentz invariance is resolved in analogy to what happens, for example, to rotational invariance in nonrelativistic Quantum Mechanics.  In this theory, rotationally-symmetric states of non-zero spin can exist, unlike in classical mechanics. These states appear the same to all observers, as they are characterized by a spherically-symmetric probability distribution. The symmetry is only broken when a direction in space is chosen for a projective measurement. That a Planck-scale fuzzyness due to Quantum Gravity might have a similar nature, has been conjectured for quite some time~\cite{Rovelli:2002vp}.

The quantum groups of interest for physics are \textit{deformations} of standard Lie group, in the sense that they depend on a set of ``noncommutativity parameters'', which, when sent to zero, make the group isomorphic to an ordinary Lie group. Under the same limit, the noncommutative homogeneous spaces associated to these quantum groups tend to some commutative homogeneous spaces. These structures are compelling candidates for a \textit{mesoscale} description of Quantum Gravity, which holds at length scales much smaller than the ones accessible with the present precision, but does not presume to capture the full complexity of physics at the Planck scale: it is supposed to represent an effective description of the first non-classical corrections to commutative spacetime.
In this spirit, one could imagine to begin with an \textit{ansatz} for the commutator between coordinate functions, which is supposed to be a coordinate-dependent antisymmetric matrix:\footnote{Such an ansatz has been explored in many papers, see for example~\cite{Doplicher:1994tu,Kempf:1996nk}.}
\begin{equation}
    [\x^\mu , \x^\nu ] :=  i \, \Theta^{\mu\nu} (\x )  \,,
\end{equation}
with the dimensions of a squared length. This matrix is supposed to be analytic in the coordinates, and to become appreciably large only when the coordinates are localized around a region of size $\ell$, where $\ell$ is an ``ultraviolet'' length scale comparable (but not necessarily identical) to the Planck length. Then, $\Theta^{\mu\nu} (\x )$ can be expanded as a power series in the coordinates ($\hat 1$ is the identity element of the algebra): 
\begin{equation}\label{CommutatorPlanckLenghExpansion}
    [\x^\mu , \x^\nu ] =  i \, \ell^2 \, \theta^{\mu\nu} \, \1 + i 
   \, \ell  \, c^{\mu\nu}{}_\rho \,  \x^\rho + i \, d^{\mu\nu}{}_{\rho\sigma} \, \x^\rho \, \x^\sigma   \,,
\end{equation}
where the coefficients $\theta^{\mu\nu}$, $c^{\mu\nu}{}_\rho$ and $d^{\mu\nu}{}_{\rho\sigma}$  are dimensionless constants, expected to be of order one.\footnote{Unless a second (or more) length scale enters the picture, because of some yet unknown physical mechanism, and one has an effective theory with two different characteristic scales. That situation could be accommodated, within the framework of Eq.~\eqref{CommutatorPlanckLenghExpansion}, by some coefficients taking values much larger or much smaller than one (\textit{i.e.} of the order of appropriate powers of the ratio between the two length scales.} Higher powers of $\x$ are excluded by a reasonable request: that, as $\ell \to 0$, the above admits a regular limit. In this same limit, the quadratic terms in $\x$ are the only ones that survive, because they do not depend at all on $\ell$, and their coefficients  $d^{\mu\nu}{}_{\rho\sigma}$ are dimensionless. These terms, \textit{q-deformation-like}~\cite{Chari}, have to be tiny in order to be compatible with observations at large scales, and will be similarly small \textit{at all scales:} the coefficients $d^{\mu\nu}{}_{\rho\sigma}$ have to be minuscule dimensionless numbers. For example, in models of deformed rotations that play a role in Loop Quantum Gravity with a cosmological constant, such terms are supposed to be of the order of the \textit{square} of the ratio between the Planck length and the Hubble scale: $\sim 10^{-122}$~\cite{Smolin:2002sz}. Such a small number might have some observable implications, in some very special regimes, but in the vast majority of situations it is bound to be highly suppressed with respect to the other terms in the expansion~\eqref{CommutatorPlanckLenghExpansion}. We can therefore safely disregard these corrections, at least in first approximation. The two remaining terms have very different dependences on the ultraviolet scale, and can be comparable to each other only at scales of the order of $\ell$. Whenever we have only access to larger scales, and we are interested in the first corrections to commutativity, the linear terms in $\x$ (sometimes dubbed ``Lie-algebraic noncommutativity''~\cite{Lukierski:2005fc}) are clearly expected to be the largest. At any rate, the ``constant'' terms $ \theta^{\mu\nu} \, \1$ represent a very mild deformation of the purely Lie-algebraic commutators, which adds very little complication, so I will keep them for the rest of the paper, also in order to be able to compare my results with those of theories that only take the $ \theta^{\mu\nu} \, \1$ term into consideration. Moreover, from now on, and unless I specify otherwise, I will use ``Planck'' units such that  $\hbar = c = \ell =1$ so the scale $\ell$ will be implicit in the definition of the unit of length.

As it turns out, not all algebras of the form~\eqref{CommutatorPlanckLenghExpansion} admit a quantum group that leaves them invariant. In this paper, I will identify 16 classes of models associated to non-isomorphic Lie algebras of noncommutative coordinates. These do not exhaust all possible real four-dimensional Lie algebras.

It is appropriate, at this point, to list three emblematic examples of quantum groups which admit a quantum homogeneous space of the form~\eqref{CommutatorPlanckLenghExpansion}. If the Lie-algebraic terms vanish, one is left with commutators between coordinates that are isomorphic to the canonical commutation relations of a particle in two dimensions, $[q^i,p_j]= i \, \hbar \, \delta^i{}_j$, $i,j =1,2$.\footnote{The real antisymmetric matrix $\theta^{\mu\nu}$ can have at most two pairs of complex-conjugate eigenvalues. If both are nonzero, there is a linear redefinition of the coordinates that transforms the matrix into the canonical symplectic form. If one eigenvalue pair is zero, then we have a two-dimensional Abelian subalgebra, and the noncommutativity is confined to two (linear combinations) of the coordinates.} This is a 4-dimensional generalization of the so-called \textit{Moyal plane,} $[\x,\hat{y}] = i \, \theta$~\cite{groenewold1946principles,moyal1949quantum,Szabo:2001kg}, and since the noncommutativity matrix is traditionally indicated with the greek letter $\theta$, this spacetime is usually called \textit{$\theta$-Minkowski,} and the corresponding quantum isometry group is called  \textit{$\theta$-Poincar\'e}~\cite{Oeckl:2000eg}.
This quantum group is defined by the following Hopf-algebraic structures:
\begin{equation}
\label{Lambda_a_coproducts_counits_antipodes}
     \begin{aligned}
&\Delta( \La^\mu{}_\nu  ) = \La^\mu{}_\rho  \otimes \La^\rho{}_\nu \,,&
&\epsilon(\La^\mu{}_\nu ) = \delta^\mu{}_\nu  \,,&
&S(\La^\mu{}_\nu ) = (\La^{-1})^\mu{}_\nu   \,,
     \\
     &\Delta( \A^\mu  ) = \La^\mu{}_\nu  \otimes  \A^\nu + \A^\mu \otimes \1 \,,&
     &\epsilon(\A^\mu ) = 0  \,,&
     &S(\A^\mu) = - (\La^{-1})^\mu{}_\nu \, \A^\nu  \,,
     \end{aligned}
\end{equation}
where $\La^\mu{}_\nu $ and $\A^\mu$ are coordinate functions on the Poincar\'e group that express the matrix elements of the Lorentz transformations and the translation parameters. The maps $\Delta$, $\epsilon$ and $S$ are the \textit{coproduct,} \textit{counit} and \textit{antipode,} which in this case codify the Lie group properties of the (standard/undeformed\footnote{In the Hopf algebra jargon, undeformed structures are usually called \textit{primitive.}}) Poincar\'e group: $\Delta$ dictates the composition law of two consecutive Poincar\'e transformations, $\epsilon$ determines the coordinates of the identity element, and $S$ gives the coordinates of the inverse of any given element. As I remarked, the rules~\eqref{Lambda_a_coproducts_counits_antipodes} are completely undeformed and independent of $\theta^{\mu\nu}$: the Lie group structure is unchanged. Where the $\theta$-Poincar\'e group differs from the standard Poincar\'e group is in the algebraic sector (the product map): the Abelian algebra $C[ISO(3,1)]$ of functions on the Poincar\'e group is replaced by an algebra $C_\theta[ISO(3,1)]$, such that:\footnote{The $\theta$-Poincar\'e group can be defined by means of a \textit{Drinfel'd Twist,}~\cite{Drinfel'd1988} and actually represents the simplest case of twisted Hopf algebra, because the carrier algebra of the twist is Abelian.}
\begin{equation}\label{Theta-Poincare}
\begin{gathered}
  [ \A^\mu , \A^\nu ]  =  i \, \theta^{\rho\sigma} \,  \left( \delta^{[\mu}{}_\rho \, \delta^{\nu]}{}_\sigma - \La^{[\mu}{}_\rho \, \La^{\nu]}{}_\sigma    \right)  
  \,,
\qquad
[ \La^{\mu}{}_\rho ,  \La^{\nu}{}_\sigma ] = 
[ \La^{\mu}{}_\rho  , \A^{\nu} ]   = 0\,,
\\
\eta_{\mu\nu}  \,   \La^\mu{}_\rho \, \La^\nu{}_\sigma 
= \eta_{\rho\sigma} 
\,, \qquad
\eta^{\rho\sigma}  \,   \La^\mu{}_\rho \, \La^\nu{}_\sigma 
= \eta^{\mu\nu} \,,
 \end{gathered}
\end{equation}
where $\eta_{\mu\nu}$ is any (constant) commutative symmetric matrix (the metric of spacetime).  It is easy to confirm that Eqs.~\eqref{Lambda_a_coproducts_counits_antipodes} together with~\eqref{Theta-Poincare} verify all of the axioms of a Hopf algebra~\cite{majid_2002}. The invariance of the $\theta$-Minkowski commutation relations (I will call the corresponding algebra $C_\theta[\mathbbm{R}^{3,1}]$):
\begin{equation}\label{Theta-Minkowski}
    [\x^\mu , \x^\nu ] =    i   \, \theta^{\mu\nu} \, \1 
\end{equation}
under this quantum group is expressed in terms of a left \textit{coaction:}
\begin{equation}\label{CoactionDef}
    (\, \cdot \,)' : C_\theta[\mathbbm{R}^{3,1}] \to C_\theta[ISO(3,1)] \otimes C_\theta[\mathbbm{R}^{3,1}] \,, \qquad
\x'^\mu = \La^\mu{}_\nu \otimes \x^\nu + \A^\mu \otimes \1 \,,
\end{equation}
which is such that 
\begin{equation}\label{CoactionInvariance}
    [\x'^\mu , \x'^\nu ] =  i  \, \theta^{\mu\nu} \, \1  \,.
\end{equation}
One can see here in what sense the noncommutativity scale is Poincar\'e-invariant: after a quantum-group change of frame, the transformed noncommutative coordinates keep closing an algebra that is identical to the original one~\eqref{Theta-Minkowski}. The uncertainty relations that this algebra codifies, and the associated limitations to the simultaneous localization of incompatible coordinates, appear exactly the same no matter how each observer is rotated, boosted or translated. The Hopf-algebra axioms (specifically, the homomorphism property of the coproduct) imply a similar notion of covariance for the coaction of the group on itself. Calling
\begin{equation}
    \La'^\mu{}_\nu = \Delta( \La^\mu{}_\nu  ) = \La^\mu{}_\rho  \otimes \La^\rho{}_\nu \,, ~~ \A'^\mu = \Delta( \A^\mu  ) = \La^\mu{}_\nu  \otimes  \A^\nu + \A^\mu \otimes \1 \,,
\end{equation}
then the commutation relations of the ``composition'' of two consecutive Poincar\'e transfomation closes the same algebra as~\eqref{Theta-Poincare}
\begin{equation}
     [ \A'^\mu , \A'^\nu ]  =  i \, \theta^{\rho\sigma} \,  \left( \delta^{[\mu}{}_\rho \, \delta^{\nu]}{}_\sigma - \La'^{[\mu}{}_\rho \, \La'^{\nu]}{}_\sigma    \right)  
  \,,
\qquad
[ \La'^{\mu}{}_\rho ,  \La'^{\nu}{}_\sigma ] = 
[ \La'^{\mu}{}_\rho  , \A'^{\nu} ]  = 0 \,.
\end{equation}
This model is appealing in its simplicity, and there is a sizeable literature of attempts at defining a QFT on it~\cite{Douglas:2001ba,Szabo:2001kg}, and even at deriving phenomenological consequences from it.  QFT predictions were originally derived using a \textit{star product}, a representation of the noncommutative product of coordinates on a space of commutative functions, in terms of an infinite series of derivatives. The Lagrangian is interpreted as a nonlocal expression involving an infinite number of derivatives of some commutative fields, and if taken at face value it would correspond to a Lorentz-violating action. Treating the commutative fields as the fundamental ones and calculating their correlation functions, one gets Lorentz-breaking results.\footnote{Exactly what happens in String-Theory-motivated models~\cite{Seiberg:1999vs}, where the $\theta^{\mu\nu}$ matrix is a tensor field with a nonzero vacuum expectation value, which provides a Lorentz-breaking background for the effective field theory.}
This conflicts with the ``quantum Poincaré invariance'' presupposed by the existence of the $\theta$-Poincar\'e group~\eqref{Theta-Poincare}, and the issue of how to properly implement ``$\theta$-Poincar\'e invariance'' in these theories has been discussed extensively~\cite{Chaichian:2004za,Wess:2003da,Koch:2004ud,Chaichian:2005yp,Bichl_2002,matlock2005noncommutative,Vitale:2023znb}. A methodology introduced in~\cite{Fiore:2007vg,Fiore:2008ta}, approached the issue of $\theta$-Poincar\'e covariance by introducing a more nuanced treatment of \textit{multilocal functions,} based on the concept of \textit{braiding}~\cite{majid_1995,Aschieri:2005zs}. In commutative QFT, the N-point functions, which completely characterize a theory, are functions mapping multiple points on the spacetime manifold to either the complex or  the real numbers. To express them in terms of the algebra of functions on the manifold $C[\mathbbm{R}^{3,1}]$, they can be viewed as elements of the tensor product of N copies of the same algebra, \textit{i.e.}  $C[\mathbbm{R}^{3,1}]^{\otimes N}$. This is legitimate, as the representation of the Poincar\'e group on spacetime points extends trivially to tensor product  representations. However, in the noncommutative context, a straightforward application of the tensor product construction fails to establish a covariant structure. In other words, assuming commutativity of coordinates among different points, as in:
\begin{equation}
    [ \x^\mu_a , \x^\nu_b ] = 0 \,, \qquad a,b \in \mathbbm{N} \,,
\end{equation}
does not maintain covariance under quantum group Poincaré transformations, if one extends the coaction~\eqref{CoactionDef} to N points as
\begin{equation}\label{PoincareCoactionNpointCoordinate}
    \x'^\mu_a = \La^\mu{}_\nu \otimes \x^\nu_a + \A^\mu \otimes \1 \,,
\end{equation}
then $    [ \x'^\mu_a , \x'^\nu_b ] \neq 0$. This is a well-known feature of the representation theory of quantum groups~\cite{majid_1995}, and it is handled by introducing a deformed (``braided'') version of the tensor product, which in the case of $\theta$-Minkowski takes the form
\begin{equation} \label{ThetaBraiding}
     [ \x^\mu_a , \x^\nu_b ] = i \, \theta^{\mu\nu} \, \1 \,, \qquad a,b \in \mathbbm{N} \,.
\end{equation}
I will call the braided tensor product algebra of $\theta$-Minkowski $C_\theta[\mathbbm{R}^{3,1}]^{\otimes_\theta N}$. The algebra~\eqref{ThetaBraiding} is covariant under~\eqref{PoincareCoactionNpointCoordinate}, and reduces to~\eqref{Theta-Minkowski} when considering the same point $a=b$. Crucially, the differences between coordinates of different points, $\x_a^\mu - \x_b ^\mu$ close an Abelian subalgebra of $C_\theta[\mathbbm{R}^{3,1}]^{\otimes_\theta N}$, so that any translation-invariant function will be commutative~\cite{Fiore:2007vg,Fiore:2008ta}. The N-point functions of the theory are therefore commutative and do not involve any additional interpretational challenge, compared to commutative QFT. In~\cite{Fiore:2007vg,Fiore:2008ta}, this construction was used to define the Wightman functions of a scalar interacting QFT, which was found to be equivalent, at least at the perturbative level, with the commutative theory.  A later approach~\cite{Nguyen:2021rsa,DimitrijevicCiric:2021jea,Giotopoulos:2021ieg,DimitrijevicCiric:2023hua,Bogdanovic:2023izt}, based on Oeckl's algebraic definition of braided QFT~\cite{Oeckl:1999zu,Oeckl:2000eg} find instead some differences, in particular beyond two-point functions.

The second-most popular noncommutative spacetime of the form~\eqref{CommutatorPlanckLenghExpansion} is called $\kappa$-Minkowski, and the associated quantum group of isometries is $\kappa$-Poincar\'e. This is actually a family of three inequivalent models. It is convenient to parametrize the whole family with four parameters $v^\mu$, $\mu = 0,\dots,3$ and write the commutation relations of the homogeneous spacetime and of the quantum group coordinates in the following way
   \begin{gather} \label{kappa-Minkowski}
    [\x^\mu,\x^\nu]= i \, (v^\mu \, \x^\nu - v^\nu \, \x^\mu ) \,,
   \\
    \begin{gathered} \label{kappa-Poincare}
    [\A^\mu,\A^\nu]= i \, (v^\mu \, \A^\nu - v^\nu \, \A^\mu ) \,,   \qquad [\La^\mu{}_\nu ,\La^\rho{}_\sigma ]=0 \,,
    \\
    [\A^\rho , \La^\mu{}_\nu ]=i \, [(\La^\mu{}_\alpha \, v^\alpha - v^\mu ) \, \La^\rho{}_\nu + (\La^\alpha{}_\nu\,\eta_{\alpha \beta}-\eta_{\nu \beta}) \,v^{\beta} \, \eta^{\mu\rho}]\,,
    \end{gathered}
    \\ \label{MetricInvariance}
    \eta_{\mu\nu}  \,   \La^\mu{}_\rho \, \La^\nu{}_\sigma 
= \eta_{\rho\sigma} 
\,, \qquad
\eta^{\rho\sigma}  \,   \La^\mu{}_\rho \, \La^\nu{}_\sigma 
= \eta^{\mu\nu} \,,
\end{gather}
where the coproduct, antipode and counit maps are still given by~\eqref{Lambda_a_coproducts_counits_antipodes}. The above commutation relations, together with~\eqref{Lambda_a_coproducts_counits_antipodes} define a Hopf algebra $C_\kappa[ISO(3,1)]$ (and a Hopf-algebra comodule $C_\kappa[\mathbbm{R}^{3,1}]$) for all choices of parameters $v^\mu$ and of (nondegenerate) metric $\eta_{\mu\nu}$.
The form of the commutation relations~\eqref{kappa-Poincare} is rather special: the commutators between translations and Lorentz matrices depends only on $\La^\mu{}_\nu$, and, since the Jacobi identities are satisfied, this can be seen as an action of $\A^\mu$ on the Hopf subalgebra closed by the Lorentz matrices (two consecutive commutators acting on $\La^\mu{}_\nu$ is the same as the action of the commutator of the $\A^\mu$'s). At the same time, the coproduct of translation~\eqref{Lambda_a_coproducts_counits_antipodes} includes a homogeneous part $\La^\mu{}_\nu \otimes \A^\nu$ which can be described in terms of a coaction of the Lorentz Hopf subalgebra on the translations (showing that this is a well-defined coaction is trivial, as the Lorentz Hopf subalgebra is primitive). The action-coaction pair I just described satisfy a set of compatibility conditions that give the complete Hopf algebra a \textit{bicrossproduct} structure, as defined by Majid~\cite{Majid:1988we,Majid:1994cy,Majid:2006xn}.

The Hopf algebras~\eqref{kappa-Poincare} for different choices of $v^\mu$  are all isomorphic to each other, as long as the Lorentzian norm of  $v^\mu$, \textit{i.e.}  $\eta_{\mu\nu}v^\mu v^\mu$, does not change sign. In other words, we have three classes of models, according to whether $v^\mu$ is timelike, spacelike or lightlike with respect to the metric $\eta_{\mu\nu}$. The timelike version, by far the most studied, was discovered in~\cite{Lukierski:1992dt,Lukierski:1991ff} (initially only in terms of the dual deformed Lie algebra, relations~\eqref{kappa-Poincare} appeared for the first time, as far as I know, in~\cite{Lukierski:1993wxa}). The spacelike version, introduced first in~\cite{Ballesteros:1995mi}, is probably the least studied, check~\cite{Blaut:2003wg,Lizzi:2020tci} for some discussion of its properties. The lightlike/null model was first introduced in~\cite{Ballesteros:1993zi,Ballesteros:1995mi,Ballesteros:1996awf}, and it turns out to be the only well-behaved one, for what concerns the goals of the present paper. In fact, it was only recently that the issue of the covariance of N-point algebras in $\kappa$-Minkowski was raised. In~\cite{Lizzi:2021rlb}, we analyzed the conditions under which a covariant braided tensor product construction could be defined on $\kappa$-Minkowski, and found that an algebra of the form
\begin{equation}\label{kappa-braiding}
 [\x_a^{\mu} , \x_b^{\nu}] 
    =  i  \,    v^{[\mu} \, (     \,  \x_a^{\nu]}  +    \x_b^{\nu]}    ) 
      + i   \, (  v^{(\mu} \, \delta^{\nu)}{}_\rho -  v_\rho \,  \eta^{\mu\nu}   ) \,  \left( \x_a^\rho - \x_b^\rho   \right)   \,,
\end{equation}
satisfies all requirements (covariance, and reducing to~\eqref{kappa-Minkowski} when $a=b$), but it only satisfies the Jacobi identities if $\eta_{\mu\nu}v^\mu v^\nu = 0$, \textit{i.e.} in the lightlike case. In~\cite{Lizzi:2021rlb} and later works~\cite{DiLuca:2022idu,Fabiano:2023xke}, we based a consistent formulation of a complex free scalar QFT on this construction, and found that, although the free theory has undeformed two-point functions, one sees differences from the commutative case whenever more than one particle is involved. In the recent~\cite{Fabiano:2023xke}, the role played in this construction by the existence of an \textit{R-matrix} was highlighted. If $v^\mu$ is lightlike, in fact, the commutation relations~\eqref{kappa-Minkowski}, \eqref{kappa-Poincare} and~\eqref{kappa-braiding} can all be written in terms of a numerical\footnote{Throughout the paper, with ``R-matrix'' I will only mean a 5-dimensional representation of an R-matrix, not a \textit{universal} one, which is  harder to find (and not really necessary for the purposes of this paper).} matrix $R^{AB}{}_{CD}$ in the following way
\begin{gather}
   \x^A \, \x^B  =   \,  R^{BA}_{CD}  \,  \x^C \, \x^D  \,,  \label{RXX_relations}
   \\
   R^{AB}_{EF} \, \hat T^E{}_C \, \hat T^F{}_D =  \hat T^B{}_G \, \hat T^A{}_H  \, R^{HG}_{CD} \,,     \label{RTT_relations}
    \\
    \x_a^A \,\x_b^B  =   R^{BA}_{CD} \,\x_b^C \, \x_a^D   \,,\label{RXY_relations}
\end{gather}
where $A=0,\dots,4$, and $\hat T^\mu{}_\nu$ is the standard 5-dimensional representation of the Poincar\'e group:
\begin{equation} \label{T_matrix_decomposition}
\hat T^{\mu}{}_\nu = \La^\mu{}_\nu \,, \qquad \hat T^\mu{}_\Q = \A^\mu \,,
    \qquad
\hat T^\Q{}_\mu = 0 \,,\qquad \hat T^\Q{}_\Q = \1 \,,
\end{equation}
and
\begin{equation}
\label{XA_decomposition}
\x^A = (\x^\mu ,\1)\,, \qquad \x_a^A = (\x_a^\mu ,\1)\,,
\end{equation}
so that the group axioms~\eqref{Lambda_a_coproducts_counits_antipodes} can be written as standard $C[GL(N)]$ rules:
\begin{equation}\label{T_matrix_coproduct_counits_antipodes}
    \Delta( \hat T^A{}_B  ) = \hat T^A{}_C  \otimes \hat T^C{}_B \,, \qquad \epsilon(\hat T^A{}_B) = \delta^A{}_B  \,, \qquad S(\hat T^A{}_B) = (\hat T^{-1})^A{}_B  \,.
\end{equation}
Eqs.~\eqref{RXX_relations}, \eqref{RTT_relations} and~\eqref{RXY_relations} are sometimes dubbed, respectively, ``RXX'', ``RTT'' and ``RXY'' relations. The so-called R-matrix $R^{AB}{}_{CD}$ is a solution of the \textit{quantum Yang--Baxter equation} (abbreviated here to QYBE, and often just called Yang--Baxter equation):
\begin{equation}\label{QYBE}
        R^{AB}_{IJ} \, 
    R^{IC}_{LK} \, 
    R^{JK}_{MN} 
    =
    R^{BC}_{JK} \, 
    R^{AK}_{IN} \, 
    R^{IJ}_{LM} \,,
\end{equation}
this implies that a non-Abelian algebra described by the relations~\eqref{RTT_relations} will satisfy a \textit{Poincar\'e--Birkhoff--Witt property}~\cite{wess1999qdeformed}, which ensures that, once a finite set of generators of the algebra are specified, the \textit{ordered} monomials in any given order are a basis for polynomials of fixed degree. Not all Hopf algebra admit a quantum R-matrix. The property of possessing an R-matrix is called \textit{quasitriangularity,} and in the case of lightlike $\kappa$-Poincar\'e an additional condition, called \textit{triangularity,}  holds: $R^{AB}_{IJ} R^{JI}_{CD} = \delta^A{}_C \, \delta^B{}_D$. The non-lightlike $\kappa$-Poincar\'e groups do not admit an R-matrix, only an \textit{r-matrix}, which, when the R-matrix exists, is the first-order approximation to it, and satisfies a linearized version of the QYBE, called (modified) classical Yang--Baxter equation, (m)CYBE in short. The absence of an R-matrix imposes severe limitations to attempts at defining a QFT on the timelike or spacelike version of $\kappa$-Poincar\'e~\cite{Arzano:2022vmh}. Notice that the $\theta$-Poincar\'e model~\eqref{Theta-Poincare} too admits an R-matrix, and can be rewritten in terms of Eqs.~\eqref{RXX_relations}, \eqref{RTT_relations} and~\eqref{RXY_relations}. Moreover, the form of the RTT relations~\eqref{RTT_relations} clarifies some of the regularities that the reader might have begun to notice, namely that the commutation relations~\eqref{Theta-Poincare} and~\eqref{kappa-Poincare} both have constant, linear and at most quadratic terms in the coordinates $\La^\mu{}_\nu$ and $\A^\mu$: this is a consequence of the fact that Eqs.~\eqref{RXY_relations} are quadratic in the components of $\hat T^A{}_B$, which, in turn, are either linear in  $\La^\mu{}_\nu$ and $\A^\mu$, or equal to the identity, see Eqs.~\eqref{T_matrix_decomposition}.

Another class of  Hopf-algebra deformations of the Poincar\'e group, admitting noncommutative homogeneous Minkowski spaces with Lie-algebraic commutators, was found long ago by Lukierski and Woronowicz~\cite{Lukierski:2005fc} (some special cases were already discussed in~\cite{Lukierski:1993hk}). This family was built using a Drin'feld twist like $\theta$-Poincar\'e, and it is parametrized by four real numbers, $\zeta^\mu$:
\begin{equation}\label{zeta-Poincare}
    \begin{gathered}
        \left[\A^{\mu },\hat{a}^{\nu }\right] = i \, \zeta ^{\nu } ( \delta _{\ \alpha }^{\mu }\hat{a}_{\beta }-\delta _{\ \beta }^{\mu }\hat{a}_{\alpha }) + \frac{i}{\kappa}\zeta^\mu( \delta _{\ \beta }^{\nu }\hat{a}_{\alpha }-\delta _{\ \alpha }^{\nu}\hat{a}_{\beta })  \,,
        \qquad [\La^\mu{}_\nu ,\La^\rho{}_\sigma ]=0 \,,
        \\
        [\A^\mu ,\La^\nu{}_\rho] = i \, \zeta ^{\lambda }\La^\mu_\lambda \, ( \eta _{\beta \rho }\, \La^\nu{}_\alpha - \eta _{\alpha \rho} \, \La^\nu{}_\beta ) + i \, \zeta ^{\mu }( \delta _{\ \beta }^{\nu}\La _{\alpha \rho }-\delta^\nu{}_\alpha \, \La _{\beta \rho }) \,,
    \end{gathered}
\end{equation}
where $\alpha$, $\beta$ are two (different) fixed components, and $\zeta^\alpha = \zeta^\beta = 0$.
The coproduct, counit and antipode maps are still undeformed~\eqref{Lambda_a_coproducts_counits_antipodes}. The noncommutative homogeneous spacetime associated to~\eqref{zeta-Poincare} has commutators
\begin{equation}
\label{zeta-Minkowski}
[\x^\mu,\x^\nu ] =  i  \, \zeta^\mu \, (\eta^{\nu\beta} \, \x^\alpha - \eta^{\nu\alpha} \, \x^\beta ) 
-
i  \, \zeta^\nu \, (\eta^{\mu\beta} \, \x^\alpha - \eta^{\mu\alpha} \, \x^\beta ) \,.
\end{equation}
Recently, a particular model with $\zeta^\mu = \delta^\mu{}_0$ and $\alpha = 1$, $\beta =2$, termed ``$\varrho$-Poincar\'e/$\varrho$-Minkowski''  has been revisited in~\cite{Lizzi:2022hcq,Fabiano:2023uhg}, its bicrossproduct structure highlighted and the relation with its definition based on a Drin'feld twist analyzed.

All the examples of quantum Poincar\'e groups I have listed have the same structure: primitive coproduct/counit/antipode maps~\eqref{Lambda_a_coproducts_counits_antipodes} codifying the Lie group structure of the Poincar\'e group,
commuting Lorentz matrices $[\La^\mu{}_\nu,\La^\rho{}_\sigma]=0$ which satisfy the standard $SO(3,1)$ constraints~\eqref{MetricInvariance}, and the remaining  commutators being polynomials of up to second order in the group coordinates $\La^\mu{}_\nu$ and $\A^\mu$. The structure is actually more specific than that: general relations of the ``RTT'' form~\eqref{RTT_relations} would
lead to all kinds of products on the right-hand side of the commutators involving the translation parameters, namely:
\begin{equation}
\begin{aligned}
[\A^\mu , \A^\nu] =& i \, \theta^{\mu\nu} \, \1 
+ i \, c^{\mu\nu}{}_\rho \, \A^\rho 
+ i \, (c_1)^{\mu\nu}{}_{\rho\sigma} \, \A^\rho \, \A^\sigma
+ i\,(c_2)^{\mu\nu\alpha}_{\beta} \, \La^\beta{}_\alpha
+
\\
& i\, (c_3)^{\mu\nu\alpha}_{\beta\gamma} \, \A^\beta \, \La^\gamma{}_\alpha
+ i \, (c_4)^{\mu\nu\alpha\gamma}_{\beta \delta} \, \La^\beta{}_\alpha  \, \La^\delta{}_\gamma \,,
\\
[\A^\mu , \La^\nu{}_\rho] =&  i \, (d_1)^{\mu\nu}_\rho  \, \1 
+ i \, (d_1)^{\mu\nu}{}_{\rho\sigma} \, \A^\sigma 
+ i \, (d_2)^{\mu\nu}{}_{\rho\sigma\lambda} \, \A^\sigma \, \A^\lambda
+ i\,(d_3)^{\mu\nu\alpha}_{\rho\beta} \, \La^\beta{}_\alpha
+
\\
& i\, (d_4)^{\mu\nu\alpha}_{\rho\beta\gamma} \, \A^\beta \, \La^\gamma{}_\alpha 
+ i \, (d_5)^{\mu\nu\alpha\gamma}_{\rho\beta \delta} \, \La^\beta{}_\alpha  \, \La^\delta{}_\gamma \,.
\end{aligned}
\end{equation}
What we see, instead, are commutation relations of the form
\begin{equation}\label{CommutatorsThatWeSee}
\begin{aligned}
[\A^\mu , \A^\nu] =& i \, \theta^{\mu\nu} \, \1 
+ i \, c^{\mu\nu}{}_\rho \, \A^\rho 
+ i\,(c_2)^{\mu\nu\alpha}_{\beta} \, \La^\beta{}_\alpha
+ i \, (c_4)^{\mu\nu\alpha\gamma}_{\beta \delta} \, \La^\beta{}_\alpha  \, \La^\delta{}_\gamma \,,
\\
[\A^\mu , \La^\nu{}_\rho] =&  i \, (d_1)^{\mu\nu}_\rho  \, \1 
+ i\,(d_3)^{\mu\nu\alpha}_{\rho\beta} \, \La^\beta{}_\alpha
+ i \, (d_5)^{\mu\nu\alpha\gamma}_{\rho\beta \delta} \, \La^\beta{}_\alpha  \, \La^\delta{}_\gamma \,,
\end{aligned}
\end{equation}
for instance,  $[\A^\mu , \A^\nu]$  depends on $\A^\mu$ up to linear terms (by the argument provided above, those terms are negligible unless we are probing Planckian distance scales), and no product between translations and Lorenz matrices appears in this commutator. Moreover, the commutator  $[\A^\mu , \La^\nu{}_\rho]$  does not depend on $\A^\mu$ at all, which makes it (if the Jacobi identities are satisfied) into a right-action of the translations on the Lorentz sector. The mixed Jacobi identities then ensure that the commutators between translations are compatible with this right action, forming a structure that generalizes slightly that of a bicrossproduct as defined in~\cite{Majid:1988we,Majid:1994cy}.\footnote{Majid's bicrossproduct construction does not account for the $\theta^{\mu\nu}$ term and the $\La^\mu{}_\nu$-dependent terms in $[\A^\mu,\A^\nu]$, as the translations are supposed to close a Lie algebra.}
The last two conditions, together with the commutativity of the $\La^\mu{}_\nu$ matrices, are responsible for creating a very specific structure: the $\La^\mu{}_\nu$ matrices 
form an Abelian subalgebra, so we can still say they are coordinates on  a commutative $GL(4)$ group manifold. The 
translations act on this manifold \textit{as vector fields.} The  commutators $[\A^\mu , \La^\nu{}_\rho]$ specify the components of the four vector fields $\A^\mu$ on $GL(4)$, and the Jacobi relations involving two translations and one Lorentz matrix impose that the Lie brackets between said vector fields is closed, \textit{i.e.} it is equal to a linear combination of $\A^\mu$. This means that we have an involutive distribution on $GL(4)$, which, by Frobenius' theorem, is associated to a foliation of $GL(4)$ by 4-dimensional leaves.
The commutator $[\A^\mu , \A^\nu]$ will constrain the components of these vector fields, because it appears in the mixed Jacobi identities I just mentioned. Finally, the additional condition~\eqref{MetricInvariance} that the  $\La^\mu{}_\nu$ matrices preserve the metric adds some further constraints on the form of the possible commutation relations: $\A^\mu$, seen as vector fields on $GL(4)$, must be tangent to the $SO(3,1)$ submanifold, and therefore they define a foliation of $SO(3,1)$. These constraints are obtained by requiring that the relations~\eqref{MetricInvariance} commute with $\A^\mu$, which restricts the coefficients of $[\A^\mu , \La^\nu{}_\rho]$.

After reviewing all of the above, a question arises naturally: what is the most general quantum Poincar\'e group that has the structure described above, and admits a braiding construction which may be used for the definition of consistent QFTs?

The purpose of this paper is to answer this question. In the next Section, I will list all of the assumptions we need to make in order to clearly delineate the class of models I am interested in. In the rest of the paper, I will proceed to classify all the possible models by imposing all the constraints that follow from my assumptions. In Sec.~\ref{SecQYBE}, I will observe how the constraints I found are equivalent to requiring the existence of a solution to the Quantum Yang--Baxter Equation with a particular ansatz for the form of the R-matrix. This ansatz will turn out to be equivalent to requiring a triangular R-matrix, which in turn is completely specified by a triangular r-matrix, solution of the Classical Yang--Baxter Equation. The triangular r-matrices on the Poincar\'e algebra have been classified long ago by Zakrzewski~\cite{Zakrzewski:1994bc,Zakrzewski_1997}: there are 16 classes up to automorphisms. This identifies 16 possible models of quantum Poincar\'e group/braided Minkowski spacetime over which we can build QFTs. The well-known $\theta$, $\kappa$ and $\varrho$-models mentioned above are all sub-cases of these 16 classes.

The triangular r-matrices found  by Zakrzewski~\cite{Zakrzewski:1994bc,Zakrzewski_1997} translate immediately into the R-matrices that I will show in the following, because the expansion of these R-matrices in terms of the r-matrices stops at first order. The main original finding of the present paper is that \textit{all} R-matrices that satisfy my set of assumptions have this form, and belong to one of the 16 classes found by Zakrzewski.

I will identify the whole class of models identified in this paper with the uppercase letter T, for ``Translations''. The quantum Poincar\'e groups will be ``T-Poincar\'e,'' and the homogeneous noncommutative Minkowski spacetimes ``T-Minkowski.'' The reason for this notation is to underline that the deformation of spacetime symmetries is entirely limited to the translation sector. In all models, the Lorentz matrices  $\La^\mu{}_\nu$ are commutative, so there is a classical $SO(3,1)$ group manifold. For instance, the infinite-dimensional representations of the algebra of coordinates on the Poincar\'e group consists of operators on the Lorentz group manifold (the $\La$ matrices are multiplicative operators), possibly in a Cartesian product with some other manifold. The translation operators close an involutive set of vector fields on the Lorentz group manifold, which defines, by Frobenius theorem, a foliation on the group. Each nondegenerate leaf of this foliation will be isomorphic to the momentum space of the theory, which will be, in general, a nonlinear manifold with a Lie group structure. Finally, the commutators between translations do not include products between $\A^\mu$ and $\La^\rho{}_\sigma$,\footnote{One model in which this does not happens was in~\cite{ballesteros2003non}. In this model, the Lorentz matrices are still commutative, but some of their components play the role of structure constants for some linear commutators of the translations. The price to pay is to have to enlarge the Poincar\'e group to the Poincar\'e-Weyl group, which includes dilatations. This model could be, in my opinion, just one example of an interesting class of models that are complementary to the ones considered here.} and consist of a Lie-algebra-type part, linear in the translations, plus a $\theta$-like part which involves the Lorentz matrices.

The choice of an uppercase latin letter, T, to name this class of models is motivated by the desire to distinguish it from the particular models it encompasses, \textit{i.e.} $\theta$, $\kappa$ and $\varrho$ so far, which will be indicated by lowercase greek letters. It is my hope that the community will adopt this notation, and, should any other particular example of the T-Minkowski spacetimes / T-Poincar\'e groups be studied in detail in the future, my suggestion is to name it after a lowercase greek letter, of which we still have plenty unused ones.

In the second paper in this series~\cite{Mercati:2024rzg}, I will develop most of the technical tools that are needed to build consistent QFTs on these 16 classes of noncommutative spacetimes, in a model-independent fashion that can be easily adapted to any particular class just by following a simple recipe.

\section{Assumptions}\label{Sec2_Assumptions}

I will assume the existence of a quantum R-matrix that describes the noncommutative product of a generic ``T-Poincar\'e'' group, which I will call $C_\ell[ISO(3,1)]$, referring to the deformation parameter $\ell$, which, in the limit $\ell \to 0$, sends the quantum group to the commutative algebra of functions on the Poincar\'e group.
The R-matrix enters the picture, initially, only through the RTT relations~\eqref{RTT_relations}, where the T-matrix is decomposed as in Eq.~\eqref{T_matrix_decomposition}, \textit{i.e.} the standard 5-dimensional representation of the Poincar\'e group.  The Lorentz matrices $\La^\mu{}_\nu$ are supposed to leave invariant a numerical metric $\eta^{\mu\nu} = \eta^{\nu\mu} \in \mathbbm{R}$, $\det \eta \neq 0$, as in Eq.~\eqref{MetricInvariance}.
$\La^\mu{}_\nu$, $\A^\mu$ and the identity $\1$ are the generators of $C_\ell[ISO(3,1)]$, and the Lie group axioms will be encoded into the primitive coproduct, counit and antipode maps of Eq.~\eqref{Lambda_a_coproducts_counits_antipodes}. 

It is easy to prove that the above maps satisfy all the axioms of a Hopf algebra, regardless of what product~\eqref{RTT_relations} one assumes, as long as the latter is associative. Imposing associativity of the product is the main challenge in defining, within this setup, a consistent Hopf algebra. Also the Lorentz-invariance of the metric, Eqs.~\eqref{MetricInvariance}, is compatible with the assumption that the maps $\Delta$ and $\epsilon$ are algebra homomorphisms, and $S$ is an algebra anti-homomorphism. However,~\eqref{MetricInvariance} is not necessarily compatible with all product rules~\eqref{RTT_relations}, and imposing their compatibility leads to further constraints on the form of Eq.~\eqref{RTT_relations}, as I will show below.

I will now list the assumptions I will make on the form of the commutation relations, which will allow me to simplify substantially the R-matrix.
\begin{itemize}
\item \textbf{Assumption I}  will be the commutativity of the Lorentz matrices:
\begin{equation}\label{Assumption_1}
    [\La^\mu{}_\nu , \La^\rho{}_\sigma] = 0 \,.
\end{equation}
Beyond what has already been said, this can also be justified on dimensional grounds: the right-hand-side of ~\eqref{Assumption_1} cannot have terms linear or quadratic in $\A^\mu$, because they would have to be multiplied by the inverse of the noncommutativity  scale $\ell$, and would not have a good commutative limit. So, only dimensionless terms depending on $\La^\mu{}_\nu$ and its square can appear, but those are of the q-deformation type, and the scaling argument outlined above applies: it makes sense to ignore them if we are interested in the first ``next-to-commutative'' corrections to QFT on Minkowski spacetime.

\item \textbf{Assumption II} will be that the commutators between translations take the form shown in Eq.~\eqref{CommutatorsThatWeSee}, which I reproduce here:
\begin{equation}\label{Assumption_2}
[\A^\mu , \A^\nu] = i \, \theta^{\mu\nu} \, \1 
+ i \, c^{\mu\nu}{}_\rho \, \A^\rho 
+ i\,(c_2)^{\mu\nu\alpha}_{\beta} \, \La^\beta{}_\alpha
+ i \, (c_4)^{\mu\nu\alpha\gamma}_{\beta \delta} \, \La^\beta{}_\alpha  \, \La^\delta{}_\gamma \,,
\end{equation}
so we exclude terms quadratic in $\A^\mu$ with the same dimensional argument as before, and, crucially, we exclude a possible term involving the product $\A^\mu \, \La^\mu{}_\nu$, which would destroy the nice structure observed in the three paradigmatic examples shown in the Introduction, of translations acting as vector fields on the Lorentz group.

\item \textbf{Assumption III:} the commutators between translations and Lorentz matrices only depends on Lorentz matrices as in Eq.~\eqref{CommutatorsThatWeSee}, reproduced here:
\begin{equation}\label{Assumption_3}
[\A^\mu , \La^\nu{}_\rho] =  i \, (d_1)^{\mu\nu}_\rho  \, \1 
+ i\,(d_3)^{\mu\nu\alpha}_{\rho\beta} \, \La^\beta{}_\alpha
+ i \, (d_5)^{\mu\nu\alpha\gamma}_{\rho\beta \delta} \, \La^\beta{}_\alpha  \, \La^\delta{}_\gamma \,.
\end{equation}
All terms depending on $\A^\mu$ can be excluded with the usual dimensional argument. The absence of those terms, in fact, is the main defining feature of the class of quantum Poincar\'e groups I am interested in: the translations $\A^\mu$ can be seen as vector fields on $SO(3,1)$, and the Jacobi identities between two translations and a Lorentz matrix will turn out to be an integrability condition on these vector fields, which ensures that they identify a foliation of $SO(3,1)$ with 4-dimensional leaves.
\end{itemize}

After imposing these three conditions on the RTT relations~\eqref{RTT_relations}, we will see that the form of the R-matrix is strongly constrained, and its free parameters reduced to 64. At this point, rather than solving right away the QYBE~\eqref{QYBE} for these 64 parameters, I will try to impose all the consistency conditions that are required to have a well-defined Hopf algebra, and the requirements that I need from a theory that I can build well-behaved QFTs on:
\begin{itemize}
\item  The Jacobi identities on all the commutators found thus far.

\item The invariance of the metric~\eqref{MetricInvariance}, \textit{i.e.} the compatibility of the commutators with $\La^\mu{}_\nu$ being a Lorentz matrix.

\item The existence of a quantum Homogeneous space, whose commutation relations are expressed in terms of RXX relations~\eqref{RXX_relations}.

\item The existence of a differential calculus and of a braiding, expressed in terms of RXY relations~\eqref{RXX_relations}. These two conditions are equivalent.

\end{itemize}
As it turns out, all of this is enough to define a class of  R-matrices that satisfy the QYBE. What is more, these R-matrices are all triangular, and are in a one-to-one correspondence  with the triangular r-matrices on the Poincar\'e group that do not involve wedge products of two Lorentz generators. These have been classified by Zakrzewski long ago~\cite{Zakrzewski:1994bc,Zakrzewski_1997}, which allows me to complete the search for all the possible models allowed by my assumptions.

\section{The ansatz on the commutators}

I will begin by imposing \textbf{Assumptions I-III}, Eqs.~\eqref{Assumption_1}, \eqref{Assumption_2} and~\eqref{Assumption_3} onto the RTT relations~\eqref{RTT_relations}.
Decomposing~\eqref{RTT_relations} into its spacetime ($\mu = 0,1,2,3$) and fifth ($A=4$) components, one gets the following relations:
\begin{align}
     \label{Eq.1}
     &R^{\Q\Q}_{\Q\Q} ~ \1  + \left( R^{\Q\Q}_{\mu\Q} +R^{\Q\Q}_{\Q\mu} \right) \A^\mu + R^{\Q\Q}_{\mu\nu} ~ \A^\mu ~ \A^\nu =    R^{\Q\Q}_{\Q\Q} ~ \1 ~,
\\     \label{Eq.2}    
    &R^{\mu\Q}_{\Q\Q} ~ \1 + \left( R^{\mu\Q}_{\nu\Q} +R^{\mu\Q}_{\Q\nu} \right) \A^\nu + R^{\mu\Q}_{\nu\rho} ~ \A^\nu ~ \A^\rho =   R^{\nu\Q}_{\Q\Q} ~ \La^\mu{}_\nu +R^{\Q\Q}_{\Q\Q} ~ \A^\mu ~,
\\     \label{Eq.3}   
    &R^{\Q\mu}_{\Q\Q} ~ \1 + \left( R^{\Q\mu}_{\nu\Q} +R^{\Q\mu}_{\Q\nu} \right) \A^\nu + R^{\Q\mu}_{\nu\rho} ~ \A^\nu ~ \A^\rho  =    ~ R^{\Q \nu}_{\Q\Q} ~ \La^\mu{}_\nu  + R^{\Q\Q}_{\Q\Q} ~ \A^\mu ~,
\\     \label{Eq.4}   
    &R^{\Q\Q}_{\nu \Q} ~  \La^\nu{}_\mu   +  R^{\Q\Q}_{\nu \rho} ~ \La^\nu{}_\mu ~ \A^\rho =    R^{\Q\Q}_{\mu\Q} ~ \1 ~,
\\     \label{Eq.5}
    &R^{\Q\Q}_{\Q\nu}  ~ \La^\nu{}_\mu + R^{\Q\Q}_{\rho\nu}  ~ \A^\rho ~ \La^\nu{}_\mu =    R^{\Q\Q}_{\Q\mu} ~ \1 ~,
\\     \label{Eq.6}
    &R^{\mu\nu}_{\Q\Q} ~ \1 + \left( R^{\mu\nu}_{\Q\rho}  + R^{\mu\nu}_{\rho\Q} \right) \A^\rho + R^{\mu\nu}_{\rho\sigma} ~ \A^\rho \, \A^\sigma  =  R^{\sigma\rho}_{\Q\Q} ~ \La^\nu{}_\rho \, \La^\mu{}_\sigma  +  R^{\Q\rho}_{\Q\Q} ~ \La^\nu{}_\rho \, \A^\mu  +  R^{\rho\Q}_{\Q\Q} ~ \A^\nu  \, \La^\mu{}_\rho  +  R^{\Q\Q}_{\Q\Q} ~ \A^\nu \, \A^\mu \,,
\\     \label{Eq.7}
    &R^{\mu \Q}_{\rho \Q}  ~ \La^\rho{}_\nu
    + R^{\mu \Q}_{\rho \sigma}  ~ \La^\rho{}_\nu ~ \A^\theta  =   R^{\rho\Q}_{\nu \Q} ~ \La^\mu{}_\rho +  R^{\Q\Q}_{\nu \Q} ~ \A^\mu  ~,
\\     \label{Eq.8}
    &R^{\mu \Q}_{\Q\rho} ~  \La^\rho{}_\nu  + R^{\mu \Q}_{\sigma\rho} ~ \A^\sigma ~  \La^\rho{}_\nu =    R^{\rho\Q}_{\Q\nu} ~ \La^\mu{}_\rho + R^{\Q\Q}_{\Q\nu} ~ \A^\mu ~,
\\     \label{Eq.9}
    &R^{\Q\mu}_{\rho \Q} ~  \La^\rho{}_\nu    + R^{\Q\mu}_{\rho \sigma} ~  \La^\rho{}_\nu   ~ \A^\sigma =  R^{\Q \rho}_{\nu \Q} ~ \La^\mu{}_\rho + R^{\Q \Q}_{\nu \Q}~ \A^\mu   ~,
\\     \label{Eq.10}
    &R^{\Q\mu}_{\Q\rho}  ~ \La^\rho{}_\nu + R^{\Q\mu}_{\sigma \rho}  ~ \A^\sigma ~ \La^\rho{}_\nu =  R^{\Q \rho}_{\Q\nu} ~ \La^\mu{}_\rho + R^{\Q\Q}_{\Q\nu} ~ \A^\mu  ~,
\\     \label{Eq.11}
    &R^{\Q\Q}_{\rho\sigma} ~ \La^\rho{}_\mu~  ~ \La^\sigma{}_\nu =    R^{\Q\Q}_{\mu\nu} ~ \1 ~,
\\     \label{Eq.12}
   & R^{\mu\nu}_{\sigma \Q} ~ \La^\sigma{}_\rho   + R^{\mu\nu}_{\sigma \theta} ~ \La^\sigma{}_\rho  ~ \A^\theta  =  
    R^{\varphi \lambda}_{\rho\Q} ~ \La^\nu{}_\lambda ~ \La^\mu{}_\varphi + R^{\Q \lambda}_{\rho\Q}  ~ \La^\nu{}_\lambda ~ \A^\mu + R^{\varphi\Q}_{\rho\Q} ~ \A^\nu ~  \La^\mu{}_\varphi + R^{\Q\Q}_{\rho\Q} ~ \A^\nu ~ \A^\mu     ~,
\\     \label{Eq.13}
    &R^{\mu\nu}_{\Q\sigma}  ~  \La^\sigma{}_\rho + R^{\mu\nu}_{\lambda\sigma} ~ \A^\lambda ~ \La^\sigma{}_\rho  = R^{\varphi\lambda}_{\Q \rho}~ \La^\nu{}_\lambda ~ \La^\mu{}_\varphi
    +R^{\Q\lambda}_{\Q \rho}~ \La^\nu{}_\lambda  ~ \A^\mu 
    +R^{\varphi\Q}_{\Q \rho} ~ \A^\nu  ~ \La^\mu{}_\varphi
    +R^{\Q\Q}_{\Q \rho} ~ \A^\nu  ~ \A^\mu  ~,
\\     \label{Eq.14}
    &R^{\mu\Q}_{\lambda\varphi} ~ \La^\lambda{}_\nu~ ~ \La^\varphi{}_\rho =    R^{\varphi\Q}_{\nu\rho} ~ \La^\mu{}_\varphi + R^{\Q\Q}_{\nu\rho} ~ \A^\mu  ~,
\\     \label{Eq.15}
    &R^{\Q\mu}_{\lambda \varphi} ~  \La^\lambda{}_\nu ~ \La^\varphi{}_\rho =  R^{\Q \varphi}_{\nu\rho} ~ \La^\mu{}_\varphi + R^{\Q \Q}_{\nu\rho} ~ \A^\mu ~,
\\      \label{Eq.16}
    &R^{\mu\nu}_{\lambda\varphi} ~ \La^\lambda{}_\rho ~ \La^\varphi{}_\sigma = R^{\varphi\lambda}_{\rho\sigma} ~ \La^\nu{}_\lambda ~ \La^\mu{}_\varphi + R^{\Q\lambda}_{\rho\sigma} ~ \La^\nu{}_\lambda  ~ \A^\mu + R^{\varphi \Q}_{\rho\sigma}        ~ \A^\nu    ~ \La^\mu{}_\varphi + R^{\Q\Q}_{\rho\sigma} ~ \A^\nu~ \A^\mu ~.
\end{align}
Imposing the form~\eqref{Assumption_1}, \eqref{Assumption_2} and~\eqref{Assumption_3} onto Equations no.~\eqref{Eq.1}, \eqref{Eq.2}, \eqref{Eq.3}, \eqref{Eq.4}, \eqref{Eq.5}, \eqref{Eq.7}, \eqref{Eq.8}, \eqref{Eq.9}, \eqref{Eq.10}, \eqref{Eq.11} and \eqref{Eq.14}, I get the following constraints on the R-matrix components:\footnote{The precise logic sequence I followed is: Eq.~\eqref{Eq.11} and~\eqref{MetricInvariance} imply $R^{\Q\Q}_{\mu\nu} \propto \eta_{\mu\nu}$. Then assumption~\eqref{Assumption_1} and Eq.~\eqref{Eq.14} imply that $R^{\mu\Q}_{\rho\sigma} = 0$ and $R^{\Q\Q}_{\mu\nu}=0$. Eq.~\eqref{Eq.15} and~\eqref{Eq.7} are solved by $R^{\Q\mu}_{\nu\rho}=0$, $R^{\Q\Q}_{\mu\Q} =0$ and $R^{\mu\Q}_{\rho\Q}=\beta \, \delta^{\mu}{}_\rho$, because the operator $(\delta^\mu{}_\alpha \, \La^\beta{}_\nu -\delta^\beta{}_\nu \, \La^\mu{}_\alpha )$ is invertible. Eq.~\eqref{Eq.4} is now automatically satisfied. Eq.~\eqref{Eq.2}  implies $R^{\mu\Q}_{\Q\nu}= (R^{\Q\Q}_{\Q\Q} - \beta )\delta^\mu{}_\nu$ and $R^{\nu\Q}_{\Q\Q}=0$. Eq.~\eqref{Eq.5} implies $R^{\Q\Q}_{\Q\mu} =0$, and Eq.~\eqref{Eq.8} implies $R^{\mu\Q}_{\Q\nu} = \gamma \, \delta^\mu{}_\nu$, which means that Eq.~\eqref{Eq.2} now reads $R^{\Q\Q}_{\Q\Q} = \beta + \gamma$. Eq.~\eqref{Eq.9}
 is soled by $R^{\Q\mu}_{\nu\Q} = \delta \, \delta^\mu{}_\nu$, and, finally, Eq.~\eqref{Eq.10} gives $R^{\Q\Q}_{\Q\mu}=0$, $R^{\Q\mu}_{\Q_\nu} = \eta \, \delta^\mu{}_\nu$. This also implies that Eq.~\eqref{Eq.1} is now automatically solved.}
\begin{equation}\label{R_matrix_components_Sol_1}
    \begin{aligned}
        & R^{\Q\Q}_{\mu\nu} =0 \,,&
        & R^{\mu\Q}_{\nu\rho}  = R^{\Q\mu}_{\nu \rho} = 0  \,,&        
        & R^{\Q\Q}_{\mu \Q} =  R^{\Q\Q}_{\Q\mu} = 0 \,,&  
        & R^{\mu\Q}_{\Q\Q} =   R^{\Q \mu}_{\Q\Q}=0 \,,&
        \\
        & R^{\mu \Q}_{\nu \Q} = \beta \, \delta^\mu{}_\nu  \,,&
        & R^{\mu \Q}_{\Q\nu} =  \gamma \, \delta^\mu{}_\nu  \,,&        
        &  R^{\Q\mu }_{\nu \Q} =  \delta \, \delta^\mu{}_\nu \,,&      
        &  R^{\Q\mu}_{\Q\nu} = \eta \, \delta^\mu{}_\nu \,,& 
        \\
        &  R^{\Q\Q}_{\Q\Q} =  \beta + \gamma \,,&       
        &  R^{\Q\Q}_{\Q\Q} = \delta +  \eta   \,,           
    \end{aligned}
\end{equation}
where $\beta$, $\gamma$, $\delta$ and $\eta$ are real constants.
Notice that solving Eqs.~\eqref{Eq.7} and~\eqref{Eq.15} required observing that, if the Lorentz matrices are commutative, the tensors $\left(   \La^\lambda{}_\nu \, \La^\varphi{}_\rho \, \delta^\mu{}_\theta- \delta^\lambda{}_\nu \,\delta^\varphi{}_\rho \, \La^\mu{}_\theta \right)$ and $\left( \delta^\mu{}_\rho  \La^\sigma{}_\nu
  -   \delta^\sigma{}_\nu  \, \La^\mu{}_\rho \right) $ are invertible, and therefore their coefficients can be put to zero.

Eq.~\eqref{Eq.16} can be solved as follows: since $R^{\mu\Q}_{\lambda\varphi}  =  R^{\Q\mu}_{\nu \rho} = R^{44}_{\mu\nu} =  0$, the equation reduces to:
\begin{equation}\label{Eq.16_2}
    R^{\mu\nu}_{\lambda\varphi} \, \La^\lambda{}_\rho \, \La^\varphi{}_\sigma =  R^{\varphi\lambda}_{\rho\sigma} \, \La^\nu{}_\lambda \, \La^\mu{}_\varphi \,.
\end{equation}
Expanding the Lorentz matrices at first order around the identity, \textit{i.e.} $\La^\mu{}_\nu = \delta^\mu{}_\nu + \eta^{\rho\mu} \, A_{\rho\nu}$, where $A_{\mu\nu}$ is a generic antisymmetric matrix, and setting the coefficients of the linearly independent components of $A_{\mu\nu}$ to zero, one gets a necessary condition for the solution of Eq.~\eqref{Eq.16_2}:
\begin{equation}
\eta^{\mu[\alpha} R^{\beta]\nu}_{\rho\sigma} + R^{\mu[\beta}_{\rho\sigma} \eta^{\alpha]\nu} = R^{\mu\nu}_{\lambda\varphi} (\eta^{\lambda[\alpha} \delta^{\beta]}{}_\rho \delta^\varphi{}_\sigma + \eta^{\varphi[\alpha} \delta^{\beta]}{}_\sigma \delta^\lambda{}_\rho ) \,,
\end{equation}
which is a system of linear equations for the components  $R^{\mu\nu}_{\rho\sigma}$, whose most general solution is ($\theta$, $\sigma$, $\kappa$ and $\lambda$ are real constants):
\begin{equation}\label{R_spacetime_components}
    R^{\mu\nu}_{\rho\sigma} = \theta \, \delta^\mu{}_\rho \delta^\nu{}_\sigma + \sigma \, \delta^\mu{}_\sigma \delta^\nu{}_\rho  + \kappa \, \eta^{\mu\nu}\eta_{\rho\sigma} + \lambda \, \varepsilon^{\mu\nu}{}_{\rho\sigma} \,.
\end{equation}
Plugging this solution back into Eq.~\eqref{Eq.16_2}, one gets:
\begin{equation}
[\La^ \mu{}_\rho , \La^\nu{}_\sigma] = \frac{\lambda}{\theta} \left( \varepsilon^{\varphi\lambda}{}_{\rho\sigma}\La^\nu{}_\lambda \, \La^\mu{}_\varphi - \varepsilon^{\mu\nu}{}_{\lambda \varphi}\La^\lambda{}_\rho \, \La^\varphi{}_\sigma \right) \,.
\end{equation}
Taking the symmetric parts of the equation in $\mu,\nu$ and in $\rho,\sigma$, one gets that the commutator $ [\La^ \mu{}_\rho , \La^\nu{}_\sigma] $ is symmetric in $\mu,\nu$ and in $\rho,\sigma$. Using this, I can rewrite the equation as 
\begin{equation}
    [\La^\mu{}_\rho , \La^\nu{}_\sigma] \propto \left( \varepsilon^{\varphi\lambda}{}_{\rho\sigma} [\La^\nu{}_\lambda , \La^\mu{}_\varphi] -\varepsilon^{\mu\nu}{}_{\lambda\varphi} [\La^\lambda{}_\rho , \La^\varphi{}_\sigma] \right) \,,
\end{equation}
Listing all the independent components of this equation (considering the aforementioned symmetries of the commutator), one gets that $[\La^\mu{}_\rho , \La^\nu{}_\sigma]  = 0$. So there are no further constraints on the remaining parameters of $R^{\mu\nu}{}_{\rho\sigma}$, \textit{i.e.} $\theta$, $\sigma$, $\kappa$ and $\lambda$.

We are left with Eqs.~\eqref{Eq.6}, \eqref{Eq.12} and~\eqref{Eq.13}, which, after replacing the conditions~\eqref{R_matrix_components_Sol_1}, take the form:
\begin{align}
        &\left( R^{\mu\nu}_{\rho\sigma} - R^{\Q\Q}_{\Q\Q} \delta^\nu{}_\rho \, \delta^\mu{}_\sigma \right) \A^\rho \, \A^\sigma +  \left( R^{\mu\nu}_{\Q\rho}  + R^{\mu\nu}_{\rho\Q} \right) \A^\rho  =   R^{\rho\sigma}_{\Q\Q} \left( \La^\nu{}_\sigma \, \La^\mu{}_\rho  -  \delta^\mu{}_\rho \, \delta^\nu{}_\sigma \right)  \,,  \label{Eq.6_2}
        \\
        &R^{\mu\nu}_{\sigma \lambda} ~ \La^\sigma{}_\rho  ~ \A^\lambda + (\eta - R^{\Q\Q}_{\Q\Q} ) \La^\nu{}_\rho  \, \A^\mu  - \beta \, \A^\nu ~  \La^\mu{}_\rho  
 =  
    R^{\varphi \lambda}_{\rho\Q} ~ \La^\nu{}_\lambda ~ \La^\mu{}_\varphi -  R^{\mu\nu}_{\sigma \Q} ~ \La^\sigma{}_\rho  \,,  \label{Eq.12_2} 
        \\
        &R^{\mu\nu}_{\sigma\lambda}  \A^\sigma \, \La^\lambda{}_\rho + ( \beta - R^{\Q\Q}_{\Q\Q} ) \A^\nu \, \La^\mu{}_\rho - \eta \, \La^\nu{}_\rho  \, \A^\mu  =    R^{\varphi\lambda}_{\Q \rho}\, \La^\nu{}_\lambda \, \La^\mu{}_\varphi  - R^{\mu\nu}_{\Q\sigma}  \,  \La^\sigma{}_\rho \,.   \label{Eq.13_2}
\end{align}
Eq.~\eqref{Eq.6_2} can be written as
\begin{equation}\label{Eq.6_2bis}
\begin{aligned}
        &\kappa \, \eta^{\mu\nu} \, \A^\rho \, \A_\rho + 
\left(\theta \, \delta^\mu{}_\rho \delta^\nu{}_\sigma + ( \sigma   - R^{\Q\Q}_{\Q\Q}) \delta^\nu{}_\rho \, \delta^\mu{}_\sigma + 
\lambda \, \varepsilon^{\mu\nu}{}_{\rho\sigma}\right) \A^\rho \, \A^\sigma +  \left( R^{\mu\nu}_{\Q\rho}  + R^{\mu\nu}_{\rho\Q} \right) \A^\rho  =
\\
&R^{\rho\sigma}_{\Q\Q} \left( \La^\nu{}_\sigma \, \La^\mu{}_\rho  -  \delta^\mu{}_\rho \, \delta^\nu{}_\sigma \right)   \,,
\end{aligned}
\end{equation}
and tracing it with the metric one gets the identities
\begin{equation}
    4 \, \kappa   + \theta \,  +  \sigma   - R^{\Q\Q}_{\Q\Q} = 0 \,, \qquad  \eta_{\mu\nu} \left( R^{\mu\nu}_{\Q\rho}  + R^{\mu\nu}_{\rho\Q} \right) = 0 \,.
\end{equation}
The antisymmetric part in $\mu$,$\nu$ of the equation can be cast in the form
\begin{equation}\label{Eq.6_3}
    \left[  \left( \theta +  2\, \kappa \right) \delta^{[\mu}{}_\rho \delta^{\nu]}{}_\sigma + 
\frac \lambda 2 \, \varepsilon^{\mu\nu}{}_{\rho\sigma} \right] [\A^\rho , \A^\sigma] = -  \left( R^{[\mu\nu]}_{\Q\rho}  + R^{[\mu\nu]}_{\rho\Q} \right) \A^\rho 
+ R^{\rho\sigma}_{\Q\Q} \left( \La^{[\nu}{}_\sigma \, \La^{\mu]}{}_\rho  -  \delta^{[\mu}{}_\rho \, \delta^{\nu]}{}_\sigma \right)  \,.
\end{equation}
The left-hand side can be seen as a linear operator on the space of antisymmetric 2-tensors of the form
\begin{equation}\label{OperatorO}
    O^{\mu\nu}{}_{\rho\sigma} =
a \, \delta^{[\mu}{}_\rho \delta^{\nu]}{}_\sigma + 
b \, \varepsilon^{\mu\nu}{}_{\rho\sigma} \,,
\end{equation}
 acting on the commutator $[\A^\rho , \A^\sigma]$. This operator has determinant $\left(a^2+4 b^2\right)^3$, which is nonzero as long as $a$ and $b$ are not both zero, and admits the inverse:
 \begin{equation}\label{OperatorOinverse}
     (O^{-1})^{\mu\nu}{}_{\rho\sigma} =
\left( \frac{a}{a^2+4 b^2} \right) \, \delta^{[\mu}{}_\rho \delta^{\nu]}{}_\sigma -  
\left( \frac{2b}{a^2+4 b^2}\right) \varepsilon^{\mu\nu}{}_{\rho\sigma} \,,
\qquad
(O^{-1})^{\mu\nu}{}_{\alpha\beta}O^{\alpha\beta}{}_{\rho\sigma} = \delta^{[\mu}{}_\rho \delta^{\nu]}{}_\sigma \,.
 \end{equation}
so, as long as $\theta + 2 \, \kappa \neq 0$ or $\lambda \neq 0$, we can invert Eq.~\eqref{Eq.6_3} into:
\begin{equation}
[\A^\alpha , \A^\beta] =
\left[
a' \, \delta^{[\alpha}{}_\mu \delta^{\beta]}{}_\nu -  
b' \varepsilon^{\alpha\beta}{}_{\mu\nu} \right] 
\bigg{[}  R^{\rho\sigma}_{\Q\Q} \left( \La^{[\nu}{}_\sigma \, \La^{\mu]}{}_\rho  -  \delta^{[\mu}{}_\rho \, \delta^{\nu]}{}_\sigma \right)  
- \left( R^{[\mu\nu]}_{\Q\rho}  + R^{[\mu\nu]}_{\rho\Q} \right) \A^\rho  \bigg{]} \,,
\end{equation}
where $a' = ( \theta +2 \kappa)[(\theta +2 \kappa )^2+\lambda ^2]^{-1}$, $b' =  \lambda[(\theta +2 \kappa )^2+\lambda ^2]^{-1}$.
As long as $\La$ belongs to the proper orthochronous subgroup of the Lorentz group, the tensor $\La^{[\mu}{}_\rho \, \La^{\nu]}{}_\sigma$ commutes with the Levi-Civita tensor with two indices raised, $\varepsilon^{\alpha\beta}{}_{\mu\nu} $:
\begin{equation}\label{CommutatorLeviCivitaLambdaLambda}
     \varepsilon^{\alpha\beta}{}_{\mu\nu}  \, \La^{[\mu}{}_\rho \, \La^{\nu]}{}_\sigma =  \varepsilon^{\mu\nu}{}_{\rho\sigma}  \, \La^{[\alpha}{}_\mu \, \La^{\beta]}{}_\nu \,,
\end{equation}
and the equation turns into
\begin{equation}\label{CommutatorAA_almostfinal}
\begin{aligned}
    [\A^\alpha , \A^\beta] =& \left[ a' \, 
R^{\mu\nu}_{\Q\Q} -  b'  \, \varepsilon^{\mu\nu}{}_{\rho\sigma}  \,
R^{\rho\sigma}_{\Q\Q}  \right]  \left( \La^{[\alpha}{}_\mu \, \La^{\beta]}{}_\nu  - \delta^{[\alpha}{}_\mu \, \delta^{\beta]}{}_\nu \right)  ] +
\\
&-\left[
a' \, \delta^{[\alpha}{}_\mu \delta^{\beta]}{}_\nu -   b' \, \varepsilon^{\alpha\beta}{}_{\mu\nu} \right]  \left( R^{[\mu\nu]}_{\Q\rho}  + R^{[\mu\nu]}_{\rho\Q} \right) \A^\rho \,.
\end{aligned}
\end{equation}
We can write the right-hand side in a compact form by introducing the following shorthands:
\begin{equation}
\left[a'  \, 
\delta^{[\alpha}{}_\mu \delta^{\beta]}{}_\nu
-  
b'  \, \varepsilon^{\alpha\beta}{}_{\mu\nu}  \right]
R^{\mu\nu}_{\Q\Q}   = - i \, \theta^{\alpha\beta} \,,
 \qquad 
 \left[a' \, \delta^{[\alpha}{}_\mu \delta^{\beta]}{}_\nu - b' \varepsilon^{\alpha\beta}{}_{\mu\nu} \right] \left( R^{[\mu\nu]}_{\Q\rho}  + R^{[\mu\nu]}_{\rho\Q} \right)
= -  i \, c^{\alpha\beta}{}_\rho \,,
\end{equation}
which can be inverted uniquely as
\begin{equation}
    \begin{gathered}
R^{[\mu\nu]}_{\Q\Q}   
= - i \,   \left( (\theta+2\,\kappa ) \, \delta^{[\mu}{}_\rho \delta^{\nu]}{}_\sigma + {\sfrac \lambda 2 } \, \varepsilon^{\mu\nu}{}_{\rho\sigma}  \right) \,  \theta^{\rho\sigma}  \,, 
\\
R^{[\mu\nu]}_{\Q\lambda}  + R^{[\mu\nu]}_{\lambda\Q} 
= - i \, \left( (\theta+2\,\kappa ) \, \delta^{[\mu}{}_\rho \delta^{\nu]}{}_\sigma + 
{\sfrac \lambda 2 } \, \varepsilon^{\mu\nu}{}_{\rho\sigma}  \right)
c^{\rho\sigma}{}_\lambda \,,
\end{gathered}
\end{equation}
and the commutation relations between translation parameters~\eqref{CommutatorAA_almostfinal} take the final form:
\begin{equation}\label{CommutatorAA_final}
    [\A^\alpha , \A^\beta] = - i \, \theta^{\mu\nu}  \, \left( \La^{[\alpha}{}_\mu \, \La^{\beta]}{}_\nu  - \delta^{[\alpha}{}_\mu \, \delta^{\beta]}{}_\nu \right)   + i \, c^{\alpha\beta}{}_\rho \,  \A^\rho \,.
\end{equation}
Now consider the symmetric part (in $\mu$, $\nu$) of Eq.~\eqref{Eq.6_2bis}. It can be written as
\begin{equation}\label{AnticummutatorAAsymmetricorder}
    \kappa \, \left( \eta^{\mu\nu} \,\eta_{\rho\sigma} 
- 2 \,  \delta^\mu{}_\rho \, \delta^\nu{}_\sigma \right)  \{ \A^{\rho} , \A^{\sigma} \}  +  \left( R^{(\mu\nu)}_{\Q\rho}  + R^{(\mu\nu)}_{\rho\Q} \right) \A^\rho  =  R^{(\rho\sigma)}_{\Q\Q} \left(\La^{(\nu}{}_\sigma \, \La^{\mu)}{}_\rho   -  \delta^{(\mu}{}_\rho \, \delta^{\nu)}{}_\sigma \right) \,,
\end{equation}
where $\{ \cdot , \cdot \}$ is the anticommutator. 
$\left( \eta^{\mu\nu} \,\eta_{\rho\sigma} 
- 2 \,  \delta^\mu{}_\rho \, \delta^\nu{}_\sigma \right)$ is an invertible operator on the space of symmetric 2-tensors. The three terms in the equation above are therefore linearly independent, and we have to set the coefficients of the two $\A$-dependent terms separately to zero:\footnote{As I said, if the  R matrix is a solution of the Quantum Yang--Baxter Equation, the algebra satisfies the Poincar\'e--Birkhoff--Witt property, which means that the ordered monomials are a basis for the algebra of functions. An expression like~\eqref{AnticummutatorAAsymmetricorder}, which is written in the symmetric ordering (thanks to the anticommutator $\{ \A^{\rho} , \A^{\sigma} \}$), can only be zero if the coefficient in front of each monomial vanishes separately.}
\begin{equation}
 \kappa = 0  \,, \qquad  R^{(\mu\nu)}_{\Q\rho}  + R^{(\mu\nu)}_{\rho\Q}  = 0  \,.
\end{equation}
The operator $( \La^{(\mu}{}_\rho \, \La^{\nu)}{}_\sigma  -  \delta^{(\mu}{}_\rho \, \delta^{\nu)}{}_\sigma )$ admits the kernel $\eta^{\rho\sigma}$,  so I can write
\begin{equation} 
  R^{\mu\nu}_{\Q\Q} =  - i \,   \left( (\theta+2\,\kappa ) \, \delta^{[\mu}{}_\rho \delta^{\nu]}{}_\sigma + 
{\sfrac \lambda 2 } \, \varepsilon^{\mu\nu}{}_{\rho\sigma}  \right) \,  \theta^{\rho\sigma}  + i \, \chi \, \eta^{\mu\nu} \,,
\end{equation}
I have then found all components of $R^{\mu\nu}_{\Q\rho}  + R^{\mu\nu}_{\rho\Q}  $:
\begin{equation}
    R^{\mu\nu}_{\Q\rho}  + R^{\mu\nu}_{\rho\Q}  =   -  i \,   \left(   \, \theta \, \delta^{[\mu}{}_\alpha \delta^{\nu]}{}_\beta + 
{\sfrac \lambda 2 }\, \varepsilon^{\mu\nu}{}_{\alpha\beta} \right) c^{\alpha\beta}{}_\rho \,.
\end{equation}

We are left with Eqs.~\eqref{Eq.12_2} and \eqref{Eq.13_2}. After replacing the expression for $R^{\mu\nu}_{\rho\sigma}$, Eq.~\eqref{R_spacetime_components}, we can take the sum between the two equations. Its symmetric part in $\mu$, $\nu$ is:
\begin{equation}
    \left( R^{(\alpha \beta)}_{\gamma\Q}  + R^{(\alpha \beta)}_{\Q \gamma}  \right)\left(  \La^{(\mu}{}_\alpha \, \La^{\nu)}{}_\beta \, \delta^\gamma{}_\rho  - \delta^{(\mu}{}_\alpha \, \delta^{\nu)}{}_\beta \, \La^\gamma{}_\rho \right) \,,
\end{equation}
which is zero, as I already showed that $R^{(\alpha \beta)}_{\gamma\Q}  + R^{(\alpha \beta)}_{\Q \gamma} = 0$. The antisymmetric part can be written as
\begin{equation}
    \left(2 \, \theta  \, \delta^{[\mu}{}_\alpha \,  \delta^{\nu]}{}_\beta    +
 \lambda \, \varepsilon^{\mu\nu}{}_{\alpha\beta} \right) [ \La^{[\alpha}{}_\rho  \, \A^{\beta]} ] 
 =
  \left(  \La^{[\mu}{}_\alpha \, \La^{\nu]}{}_\beta \, \delta^\gamma{}_\rho  - \delta^{[\mu}{}_\alpha \, \delta^{\nu]}{}_\beta \, \La^\gamma{}_\rho \right)   \left( R^{[\alpha \beta]}_{\gamma\Q}  + R^{[\alpha \beta]}_{\Q \gamma}  \right) \,,
\end{equation}
where I reordered some terms using the commutativity of Lorentz matrices. Like before, we can invert the operator acting on the commutator on the left-hand-side, and write the equation as
\begin{equation}
    [ \La^{[\alpha}{}_\rho  \, \A^{\beta]} ] 
 =-  \frac{i}{2}
   (O^{-1})^{\alpha\beta}{}_{\mu\nu}   \left(  \La^{[\mu}{}_\eta \, \La^{\nu]}{}_\theta \, \delta^\gamma{}_\rho  - \delta^{[\mu}{}_\eta \, \delta^{\nu]}{}_\theta \, \La^\gamma{}_\rho \right)  O^{\eta\theta}{}_{\tau\kappa} c^{\tau\kappa}{}_\gamma \,,
\end{equation}
where $O^{\mu\nu}{}_{\rho\sigma}$ and its inverse are given by Eq.~\eqref{OperatorO} and~\eqref{OperatorOinverse}, with $a=2 \, \theta$ and $b = \lambda$. We can use again Eq.~\eqref{CommutatorLeviCivitaLambdaLambda} to prove that the tensor $\La^{[\mu}{}_\eta \, \La^{\nu]}{}_\theta \, \delta^\gamma{}_\rho  - \delta^{[\mu}{}_\eta \, \delta^{\nu]}{}_\theta \, \La^\gamma{}_\rho$ commutes with the operator $O^{\mu\nu}{}_{\rho\sigma}$, which can therefore pass to the left side of it and cancel out with its inverse. We are left with the following equation:
\begin{equation}
    [ \La^{[\mu}{}_\rho  \, \A^{\nu]} ]   =
  -  \frac{i}{2} \,   c^{\alpha\beta}{}_\gamma
\left(  \La^{[\mu}{}_\alpha \, \La^{\nu]}{}_\beta \, \delta^\gamma{}_\rho  - \delta^{[\mu}{}_\alpha \, \delta^{\nu]}{}_\beta \, \La^\gamma{}_\rho \right) \,.
\end{equation}

The last equation to use is the difference between Eq.~\eqref{Eq.12_2} and \eqref{Eq.13_2}:
\begin{equation}\label{LastEqRTT}
\begin{gathered}
        2 \, \theta  \, \{ \La^{[\mu}{}_\rho   , \A^{\nu]} \}
 +  2 \,\eta \, \La^\nu{}_\rho  \, \A^\mu   
 - 2 \,\beta \, \A^\nu \, \La^\mu{}_\rho  
+ \lambda \, \varepsilon^{\mu\nu}{}_{\alpha\beta} \left(  \La^\alpha{}_\rho  \, \A^\beta  - \A^\alpha \, \La^\beta{}_\rho \right)   = 
\\
\left( R^{\alpha \beta}_{\gamma\Q}  - R^{\alpha \beta}_{\Q \gamma}  \right)\left(  \La^\mu{}_\alpha \, \La^\nu{}_\beta \, \delta^\gamma{}_\rho  - \delta^\mu{}_\alpha \, \delta^\nu{}_\beta \, \La^\gamma{}_\rho \right) \,.
\end{gathered}
\end{equation}
\textbf{Assumption III} implies that the RTT equations have to reduce to the form~\eqref{Assumption_3}. Then, if we manage to rewrite this last equation in terms of commutators and anticommutators of $\La^\mu{}_\nu$ and $\A^\rho$, the coefficients of the anticommutators must vanish, while the coefficients of the commutators will fix the form of the free parameters in Eq.~\eqref{Assumption_3}.
Taking the symmetric part in $\mu$, $\nu$ of Eq.~\eqref{LastEqRTT}:
\begin{equation}\label{LastEqRTT_symm_part}
    (\eta + \beta)\, [\La^{(\mu}{}_\rho  , \A^{\nu)} ] 
 + (\eta - \beta) \, \{ \La^{(\mu}{}_\rho , \A^{\nu)} \} 
 =    \left( R^{(\alpha \beta)}_{\gamma\Q}  - R^{(\alpha \beta)}_{\Q \gamma}  \right)\left(  \La^{(\mu}{}_\alpha \, \La^{\nu)}{}_\beta \, \delta^\gamma{}_\rho  - \delta^{(\mu}{}_\alpha \, \delta^{\nu)}{}_\beta \, \La^\gamma{}_\rho \right) \,,
\end{equation}
I get an equation that reduces to the form~\eqref{Assumption_3} only if:\footnote{To further clarify this step, the logic is the following: Eq.~\eqref{LastEqRTT_symm_part} consists of a commutator term, an anticommutator term and a term that depends only on $\La^\mu{}_\nu$. \textbf{Assumption III} dictates that the commutator term is itself a function of $\La^\mu{}_\nu$, therefore we are left with an equation of the same form as~\eqref{AnticummutatorAAsymmetricorder}, involving a symmetrically-ordered term and a term that depends only on $\La^\mu{}_\nu$. The Poincar\'e--Birkhoff--Witt property implies that the two terms have to vanish separately.}
\begin{equation}
\eta = \beta \,, \qquad
 [\La^{(\mu}{}_\rho  , \A^{\nu)} ]  
 =  \frac{1}{2 \, \beta}   \left( R^{(\alpha \beta)}_{\gamma\Q}  - R^{(\alpha \beta)}_{\Q \gamma}  \right)\left(  \La^{(\mu}{}_\alpha \, \La^{\nu)}{}_\beta \, \delta^\gamma{}_\rho  - \delta^{(\mu}{}_\alpha \, \delta^{\nu)}{}_\beta \, \La^\gamma{}_\rho \right) \,.
\end{equation}
The antisymmetric part in $\mu$ $\nu$ of Eq.~\eqref{Assumption_3} is (using also $\eta = \beta$):
\begin{equation}
    \left(  2(  \theta  -  \beta ) \delta^{[\mu}{}_\alpha  \, \delta^{\nu]}{}_\beta 
 + \lambda \, \varepsilon^{\mu\nu}{}_{\alpha\beta}  \right)\{ \La^{[\alpha}{}_\rho  , \A^{\beta]} \}
 =    \left( R^{[\alpha \beta]}_{\gamma\Q}  - R^{[\alpha \beta]}_{\Q \gamma}  \right)\left(  \La^{[\mu}{}_\alpha \, \La^{\nu]}{}_\beta \, \delta^\gamma{}_\rho  - \delta^{[\mu}{}_\alpha \, \delta^{\nu]}{}_\beta \, \La^\gamma{}_\rho \right) \,,
\end{equation}
which, by the Poincar\'e--Birkhoff--Witt property, since the operator $(  \La^{[\mu}{}_\alpha \, \La^{\nu]}{}_\beta \, \delta^\gamma{}_\rho  - \delta^{[\mu}{}_\alpha \, \delta^{\nu]}{}_\beta \, \La^\gamma{}_\rho )$ is invertible, is solved by:
\begin{equation}
\beta = \theta \,, \qquad \lambda = 0 \,, \qquad R^{[\alpha \beta]}_{\gamma\Q}  = R^{[\alpha \beta]}_{\Q \gamma} \,.
\end{equation}
We then have all the components of the commutator between Lorentz matrices and translations:
\begin{equation}
[ \La^{\mu}{}_\rho  , \A^{\nu} ]   =
  i \,  f^{\alpha\beta}{}_\gamma
\left(  \La^{\mu}{}_\alpha \, \La^{\nu}{}_\beta \, \delta^\gamma{}_\rho  - \delta^{\mu}{}_\alpha \, \delta^{\nu}{}_\beta \, \La^\gamma{}_\rho \right)  \,,
\end{equation}
where
\begin{equation}
 f^{\alpha\beta}{}_\gamma
 =
\frac{1}{2} \, b^{(\alpha\beta)}{}_\gamma - \frac 1 2 \,  c^{[\alpha\beta]}{}_\gamma \,,
\end{equation}
and where I repackaged some components of the R-matrix into the following object, symmetric in its two upper indices:
\begin{equation}
b^{(\alpha\beta)}{}_\gamma = - \frac{i}{\theta}  \left( R^{(\alpha \beta)}_{\gamma\Q}  - R^{(\alpha \beta)}_{\Q \gamma}  \right) \,.
\end{equation}

It is time to recap everything we have learned so far about the R-matrix:
\begin{equation}\label{R_matrix_components_Sol_2}
\begin{gathered}
    \begin{aligned}
        & R^{\Q\Q}_{\mu\nu} =0 \,,&
        & R^{\mu\Q}_{\nu\rho}  = R^{\Q\mu}_{\nu \rho} = 0  \,,&        
        & R^{\Q\Q}_{\mu \Q} =  R^{\Q\Q}_{\Q\mu} = 0 \,,&  
        & R^{\mu\Q}_{\Q\Q} =   R^{\Q \mu}_{\Q\Q}=0 \,,&
        \\
        & R^{\mu \Q}_{\nu \Q} = \theta \, \delta^\mu{}_\nu  \,,&
        & R^{\mu \Q}_{\Q\nu} =  \sigma \, \delta^\mu{}_\nu  \,,&        
        &  R^{\Q\mu }_{\nu \Q} =  \sigma \, \delta^\mu{}_\nu \,,&      
        &  R^{\Q\mu}_{\Q\nu} = \theta \, \delta^\mu{}_\nu \,,& 
    \end{aligned}
    \\
         R^{\Q\Q}_{\Q\Q} = \theta + \sigma \,, 
         \qquad 
         R^{\mu\nu}_{\rho\sigma} = \theta \, \delta^\mu{}_\rho \delta^\nu{}_\sigma + \sigma \, \delta^\mu{}_\sigma \delta^\nu{}_\rho      \,,
    \\
          R^{\mu \nu}_{\rho\Q}  = i \, {\sfrac \theta 2} \left(  b^{\mu\nu}{}_\rho - c^{\mu\nu}{}_\rho \right) \,,
          \qquad
          R^{\mu \nu}_{\Q\rho}  = - i \, {\sfrac \theta 2} \left(  b^{\mu\nu}{}_\rho + c^{\mu\nu}{}_\rho \right) \,,
          \\
           R^{\mu\nu}_{\Q\Q} =  - i \, \theta \, \theta^{\mu\nu}  + i \, \chi \, \eta^{\mu\nu} 
\end{gathered}
\end{equation}
which is parametrized by the two real parameters $\theta$ and $\sigma$, plus the 24 independent components of $c^{\mu\nu}{}_\rho$, and the 40 of $b^{\mu\nu}{}_\rho $. The corresponding commutation relations of the Poincar\'e group coordinates are
\begin{equation}\label{QuantumPoincareGroupCommutators}
\begin{gathered}
  [ \A^\mu , \A^\nu ]  =  i \, \theta^{\rho\sigma} \,  \left( \delta^{[\mu}{}_\rho \, \delta^{\nu]}{}_\sigma - \La^{[\mu}{}_\rho \, \La^{\nu]}{}_\sigma    \right)  
  + i \, c^{\mu\nu}{}_\rho \, \A^\rho  \,,
\qquad
[ \La^{\mu}{}_\rho ,  \La^{\nu}{}_\sigma ] = 0   \,,
\\
[ \La^{\mu}{}_\rho  , \A^{\nu} ]   =
  i \,  f^{\alpha\beta}{}_\gamma
\left(  \La^{\mu}{}_\alpha \, \La^{\nu}{}_\beta \, \delta^\gamma{}_\rho  - \delta^{\mu}{}_\alpha \, \delta^{\nu}{}_\beta \, \La^\gamma{}_\rho \right) 
 \,.
 \end{gathered}
\end{equation}

\section{Further constraints on the R-matrix}\label{Sec4_FurtherConstraints}

\subsection{Jacobi identities and metric invariance}\label{Subsec41_JacobiMetric}

The 64-parameter algebra~\eqref{QuantumPoincareGroupCommutators} does not automatically define the multiplication of a quantum group. First of all, one needs to make sure that the coproduct, counit and antipode maps~\eqref{Lambda_a_coproducts_counits_antipodes} are compatible with the commutation relatios: the first two need to be homomorphisms, and the third an antihomomorphism for the multiplication. As I remarked in Sec.~\ref{Sec2_Assumptions}, primitive group-like coproduct, counit and antipode maps automatically satisfy this requirement, as long as the commutation relations take the RTT form~\eqref{RTT_relations}. We can double-check that the commutators~\eqref{QuantumPoincareGroupCommutators} verify this requirement by direct calculation:
\begin{equation}
    [ \Delta (\La^{\mu}{}_\rho)  , \Delta (\A^{\nu}) ]  = \Delta \left( [ \La^{\mu}{}_\rho  , \A^{\nu} ]  \right) \,, \qquad 
[ \Delta (\A^\mu )  , \Delta (\A^\nu) ] = \Delta \left( [ \A^\mu , \A^\nu ] \right)\,,
\end{equation}
\begin{equation}
    [ \epsilon (\La^{\mu}{}_\rho)  , \epsilon (\A^{\nu}) ]  = \epsilon \left( [ \La^{\mu}{}_\rho  , \A^{\nu} ]  \right) \,, \qquad 
[ \epsilon (\A^\mu )  , \epsilon (\A^\nu) ] = \epsilon \left( [ \A^\mu , \A^\nu ] \right)\,,
\end{equation}
while for the antipode a little more work is required: one has to first prove that the commutation relations~\eqref{QuantumPoincareGroupCommutators}, together with the left- and right-inverse rule $(\La^{-1})^\mu{}_\rho \, \La^\rho{}_\nu = \La^\mu{}_\rho \, (\La^{-1})^\rho{}_\nu = \delta^\mu{}_\nu$, imply that the commutators of translations with the inverse Lorentz matrix are:
\begin{equation}
 [(\La^{-1})^\mu{}_\lambda  , \A^\nu]  
 = -   i \,  f^{\alpha\beta}{}_\gamma \, 
\left(  \delta^\mu{}_\alpha \, \La^{\nu}{}_\beta \, (\La^{-1})^\gamma{}_\lambda - (\La^{-1})^\mu{}_\alpha \, \delta^{\nu}{}_\beta \, \delta^\gamma{}_\lambda \right) \,,
\end{equation}
and then, recalling that the antipode is an antihomomorphism: 
\begin{equation}
    [ S (\A^{\nu}) ,  S (\La^{\mu}{}_\rho)  ]  = S \left( [ \La^{\mu}{}_\rho  , \A^{\nu} ]  \right) \,, \qquad 
[  S (\A^\nu) , S (\A^\mu ) ] = S \left( [ \A^\mu , \A^\nu ] \right)\,.
\end{equation}
One should also check that the metric covariance relations~\eqref{MetricInvariance} are compatible with the three maps $\Delta$, $\epsilon$ and $S$, but this is guaranteed because the $\La^\mu{}_\nu$ matrices are commutative, and Eq.~\eqref{MetricInvariance} is undeformed.

What is not guaranteed, until we have found the an R-matrix that satisfies the QYBE, is that the product defined by~\eqref{QuantumPoincareGroupCommutators} is associative. For this to happen, the Jacobi identities need to be verified. These identities would be automatically satisfied if we had a solution of the QYBE, however, the equation is rather complex, having $5^6 = 15625$ components, each of which has, in principle, $5^3 = 125$ monomials. Even decomposing the equation and each monomial into spacetime and $4$ components, one still has $2^6 = 64$ components, each of which with potentially $2^3=8$ different terms. Imposing the Jacobi identities on the commutators~\eqref{QuantumPoincareGroupCommutators} is much more feasible, as I shall illustrate now.

There are only two types of nontrivial Jacobi identities. Those involving three Lorentz matrices are trivial, because they close an Abelian subalgebra. The ones involving two $\La$ and one translation are trivial too, because once a commutator between $\La$ and $\A$ is calculated, the result only depends on $\La$ and is therefore guaranteed to commute with the remaining $\La$. The two remaining ones, between two translations and one $\La$, and the one between three translations, are nontrivial.
Consider the latter first:
\begin{equation}
    [[ \A ^\mu , \A^\nu] , \A^\rho] + [[ \A ^\rho , \A^\mu] , \A^\nu] + [[ \A ^\nu , \A^\rho] , \A^\mu]= 0 \,,
\end{equation}
a simple calculation reveals that the equation above is solved by
\begin{equation}   \label{JacobiIdentitiesTranslationAlgebra}
     \begin{aligned}
        \left( c^{\mu\nu}{}_\lambda \, c^{\lambda\rho}{}_\sigma - 
        \, c^{\mu\rho}{}_\lambda c^{\lambda\nu}{}_\sigma 
        - c^{\nu\rho}{}_\lambda \, c^{\mu\lambda}{}_\sigma \right)
        + 
        \left(  c^{\mu\nu}{}_\lambda \, \theta^{\rho \lambda}   + c^{\rho\mu}{}_\lambda \, \theta^{\nu \lambda} + c^{\nu\rho}{}_\lambda \, \theta^{\mu \lambda} \right)
        = 0\,.
     \end{aligned}
\end{equation}
The first of the two brackets is just the Jacobi identities for the structure constants $c^{\mu\nu}{}_\rho$ of a Lie algebra. The second are the consistency conditions for an extension of that Lie algebra by a single central generator. These state that the matrix $\theta^{\mu\nu}$, seen as a bilinear map on the Lie algebra, is a 2-cocycle.

Now consider the Jacobi identities between one Lorentz matrix and two translations:
\begin{equation}
[[\La^\mu{}_\rho , \A^\nu],\A^\sigma]
-
[[\La^\mu{}_\rho,\A^\sigma],\A^\nu]
+
[[\A^\nu,\A^\sigma],\La^\mu{}_\rho] = 0
~~ \Rightarrow ~~ 
[[\La^\mu{}_\rho , \A^{[\nu}],\A^{\sigma]}] -
{\sfrac i 2} \, c^{\nu\sigma}{}_\lambda [\La^\mu{}_\rho,\A^\lambda] = 0 \,,
\end{equation}
looking at the structure of the above equation, as well as the commutation relation~\eqref{QuantumPoincareGroupCommutators} between $\La$ and $\A$, a familiar structure appears: the translations $\A^\mu$ act like four vector fields on the $GL(4)$ manifold that $\La^\mu{}_\nu$ coordinatize, and the Jacobi identities impose that these vector fields close an algebra under Lie brackets, with structure constants ${\sfrac 1 2} c^{\mu\nu}_{\rho}$. In other words, the $\A^\mu$ vector fields close an integrable/involutive system, which identifies a 4-dimensional foliation of $GL(4)$~\cite{lee2003introduction}.\footnote{Notice that, at this stage, I have to consider $\La^\mu{}_\nu$ as coordinates on $GL(4)$ rather than $SO(3,1)$. If we imposed that these coordinates are redundant and they satisfy the constraints~\eqref{MetricInvariance}, then further conditions would be imposed on $\A^\mu$ for them to be vector fields on $SO(3,1)$, which close an integrable system. This will be done shortly.} Straightforward, albeit tedious, calculations show that
\begin{equation}
    \begin{gathered}
   [[\La^\mu{}_\rho , \A^{[\nu}],\A^{\sigma]}] -
{\sfrac i 2} \, c^{\nu\sigma}{}_\lambda [\La^\mu{}_\rho,\A^\lambda] =
\\
\left(     f^{\varphi [\kappa}{}_\psi  \, f^{\psi |\lambda]}{}_\theta  - {\sfrac 1 2} \,  c^{\lambda \kappa}{}_\psi  \, f^{\varphi \psi}{}_\theta \right) \left( \delta^\theta{}_\rho  \, \La^{\mu}{}_\varphi  \, \La^{[\sigma}{}_\kappa \, \La^{\nu]}{}_\lambda 
  -   \delta^{\mu}{}_\varphi  \, \delta^{[\sigma}{}_\kappa \, \delta^{\nu]}{}_\lambda   \, \La^\theta{}_\rho \right) \,.
\end{gathered} 
\end{equation}
the tensor $ ( \delta^\theta{}_\rho  \, \La^{\mu}{}_\varphi  \, \La^{[\sigma}{}_\kappa \, \La^{\nu]}{}_\lambda 
  -   \delta^{\mu}{}_\varphi  \, \delta^{[\sigma}{}_\kappa \, \delta^{\nu]}{}_\lambda   \, \La^\theta{}_\rho ) $ has a nontrivial kernel:
  \begin{equation}\label{W-matrix_definition}
    W^{\mu\nu\rho}_\sigma = \tau \, \delta^{[\mu}{}_\sigma \, \eta^{\nu] \rho} + \xi \, \varepsilon^{\mu\nu\rho}{}_\sigma 
    \qquad \Rightarrow 
    \qquad
    W^{\varphi\kappa\lambda}_\theta \, \left( \delta^\theta{}_\rho  \, \La^{\mu}{}_\varphi  \, \La^{[\sigma}{}_\kappa \, \La^{\nu]}{}_\lambda 
  -   \delta^{\mu}{}_\varphi  \, \delta^{[\sigma}{}_\kappa \, \delta^{\nu]}{}_\lambda   \, \La^\theta{}_\rho \right)  = 0\,,
  \end{equation}
  therefore, the most general solution of the above equation is:
\begin{equation}\label{ModifiedJacobiIdentities_for_f}
     f^{\alpha\beta}{}_\eta  \, f^{\eta\gamma}{}_\epsilon -      f^{\alpha\gamma}{}_\eta  \, f^{\eta\beta}{}_\epsilon 
    = - c^{\beta\gamma}{}_\delta \, f^{\alpha\delta}{}_\epsilon + W^{\alpha\beta\gamma}_\epsilon \,,
\end{equation}
for $\tau$, $\xi$ arbitrary parameters.
This equation admits a simple interpretation: consider the following four matrices:
\begin{equation}
    (K^\mu)^\alpha{}_\beta = f^{\alpha\mu}{}_\beta \,.
\end{equation}
Eq.~\eqref{ModifiedJacobiIdentities_for_f} implies that their commutators are
\begin{equation}\label{Modified_commutator_K_matrices}
    [K^\mu , K^\nu] = - c^{\mu\nu}{}_\rho \, K^\rho + W^{\mu\nu} \,, 
\end{equation}
where $(W^{\mu\nu})^\alpha{}_\beta = W^{\alpha\mu\nu}_\beta$. If $\tau = \xi =0$, and therefore $W^{\mu\nu} = 0$, then the $K^\mu$ matrices form a (not necessarily faithful) representation of the Lie algebra with structure constants $c^{\mu\nu}{}_\rho$.

Before going further, I have to discuss the compatibility conditions between the multiplication~\eqref{QuantumPoincareGroupCommutators} and the invariance of the metric~\eqref{MetricInvariance}. Both sides of the two Eqs.~\eqref{MetricInvariance} commute with Lorentz matrices, but the left-hand-sides do not necessarily commute with translations, therefore we have to impose that it does. Let us begin with the left equation:
\begin{equation}\label{ConditionCommutativitywithLambdaLambdaEta}
    \begin{aligned}
&[\eta^{\rho\sigma} \La^\alpha{}_\rho \La^\beta{}_\sigma  , \A^\nu] 
  = 2 \, \eta^{\rho\sigma} \La^{(\alpha}{}_\sigma [\La^{\beta)}{}_\rho  , \A^\nu] =
2 \, i \,  \eta^{\theta (\sigma}  \,  f^{\varphi ) \lambda }{}_\theta   \left(  \La^{(\alpha}{}_\sigma \La^{\beta)}{}_\varphi \, \La^\nu{}_\lambda     -    \delta^{(\alpha}{}_\sigma \delta^{\beta)}{}_\varphi \, \delta^\nu{}_\lambda  \right) =0  \,,
    \end{aligned}
\end{equation}
expanding the Lorentz matrices in the right-hand side as a power series of an antisymmetric matrix (with one index raised with the Minkowski metric), and imposing that the coefficients of the first order of the expansion vanish, one gets the following \textit{necessary} condition for the vanishing of the whole expression:
 \begin{equation}\label{K_LorentAlgebraMatrices}
     \eta^{\theta (\sigma}  \,  f^{\varphi ) \lambda }{}_\theta = \eta^{\theta (\sigma}  \, (K^\lambda)^{\varphi )}{}_\theta = 0 \,.
 \end{equation}
It is immediate to verify that Eq.~\eqref{K_LorentAlgebraMatrices} is a solution of the original equation~\eqref{ConditionCommutativitywithLambdaLambdaEta}, and therefore~\eqref{K_LorentAlgebraMatrices} is also a \textit{sufficient} condition for Eq.~\eqref{ConditionCommutativitywithLambdaLambdaEta} to be satisfied, and we have thus found the general solution.
 
Eq.~\eqref{K_LorentAlgebraMatrices} means that the matrices $K^\mu$ defined above are \textit{Lorentz algebra matrices,} because it imposes that, once one of their matrix indices is raised with the Minkowski metric, they are antisymmetric. We can parametrize all possible Lorentz algebra generators with antisymmetric matrices with one index raised (or lowered) with the metric, so we can solve Eq.~\eqref{K_LorentAlgebraMatrices} as
  \begin{equation} \label{f_parametrization_Lorentz_matrices}
f^{\mu\nu}{}_\rho = \eta^{\mu\sigma} \omega^\nu_{\sigma\rho} 
  \,,
 \end{equation}
 where $\omega^\nu_{\sigma\rho} = - \omega^\nu_{\rho\sigma}$ are four independent antisymmetric matrices. Having fixed the $f^{\mu\nu}{}_\rho$ constants, we can write the  $b^{\mu\nu}{}_\rho $ constants and the Lie algebra structure constants $c^{\mu\nu}{}_\rho $, which are, respectively,  the symmetric and antisymmetric parts of $f^{\mu\nu}{}_\rho $, in terms of the $\omega^\nu_{\sigma\rho}$ matrices:
 \begin{equation}
  b^{\mu\nu}{}_\rho =     f^{\mu\nu}{}_\rho  +  f^{\nu\mu}{}_\rho  =  \eta^{\mu\sigma} \omega^\nu_{\sigma\rho}  + \eta^{\nu\sigma} \omega^\mu_{\sigma\rho} \,,
\qquad
  c^{\mu\nu}{}_\rho =    - f^{\mu\nu}{}_\rho  +  f^{\nu\mu}{}_\rho  =  - \eta^{\mu\sigma} \omega^\nu_{\sigma\rho}  + \eta^{\nu\sigma} \omega^\mu_{\sigma\rho} \,.
\end{equation}
Alternatively, if we want to define our Hopf algebra in terms of the structure constants of a Lie algebra $c^{\mu\nu}{}_\rho $ (which will determine the commutators between spacetime coordinates, see below) and the metric $\eta_{\mu\nu}$, the equation on the right can be inverted as follows:
\begin{equation}
 \omega^\mu_{\rho\sigma} = - {\sfrac 1 2} \left(
       \eta_{\sigma\lambda}  c^{\mu\lambda}{}_\rho
       -
       \eta_{\rho\lambda}  c^{\mu\lambda}{}_\sigma
   +  \eta^{\mu\nu}\eta_{\rho\lambda} \eta_{\sigma\kappa}  c^{\lambda\kappa}{}_\nu
       \right) 
\,.
\end{equation}
 The other metric-invariance relation in~\eqref{MetricInvariance} is now automatically satisfied. In fact:
\begin{equation}
\eta_{\alpha\beta} \, [\La^\alpha{}_\rho \La^\beta{}_\sigma  , \A^\nu] 
=
2 \, i \, f^{\varphi \lambda}{}_{(\theta}  \, \eta_{\alpha)\varphi} \,  \left( \La^\nu{}_\lambda  \, \delta^{(\alpha}{}_{\sigma} \, \delta^{\theta)}{}_{\rho}   -   \delta^\nu{}_\lambda \, \La^{(\alpha}{}_{\sigma}  \, \La^{\theta)}{}_{\rho}  \right) \,,
\end{equation}
and, using~\eqref{f_parametrization_Lorentz_matrices},  the expression above vanishes, because
\begin{equation}
    f^{\varphi \lambda}{}_{(\theta}  \, \eta_{\alpha)\varphi}
    = \eta^{\varphi\sigma} \omega^\lambda_{\sigma(\theta}  \, \eta_{\alpha)\varphi} = \omega^\lambda_{(\alpha\theta)}  = 0 \,.
\end{equation}

We now have a Hopf algebra for any solution of Eqs.~\eqref{JacobiIdentitiesTranslationAlgebra} and~\eqref{Modified_commutator_K_matrices}, subject to the conditions~\eqref{K_LorentAlgebraMatrices}.  If we can find a (not necessarily nondegenerate) representation of the Lie algebra with structure constants~$c^{\mu\nu}{}_\rho $ \textit{in terms of Lorentz algebra generators,} then we have a Hopf algebra. This does not necessarily mean that the Lie algebra with structure constants~$c^{\mu\nu}{}_\rho $ is a subalgebra of the Lorentz algebra. In fact, one can for example take, for some of the matrices $K^\mu$, copies of the same Lorentz algebra generators, or the zero matrix. Then some of the components of Eqs.~\eqref{K_LorentAlgebraMatrices} can be trivially satisfied, for example having the right-hand-side vanish.

To make this concrete, let us see the examples of $\kappa$-Poincar\'e and  $\varrho$-Poincar\'e written in this language.
The structure constants of the $\kappa$-Poincar\'e family can be written as
\begin{equation}
\begin{gathered}
     f^{\mu\nu}{}_\rho =  v_\rho \,  \eta^{\mu\nu} - v^{\mu} \, \delta^\nu{}_\rho   \,  \,,
    \qquad b^{\mu\nu}{}_\rho =  2 \, ( v_\rho \,  \eta^{\mu\nu} - v^{(\mu} \, \delta^{\nu)}{}_\rho   )  \,, 
\qquad
 c^{\mu\nu}{}_\rho =  2 \, v^{[\mu} \, \delta^{\nu]}{}_\rho   \,,
\end{gathered}
\end{equation}
and the corresponding $K^\mu$ matrices take the following form:
\begin{equation}
\begin{aligned}
&K^0 =
-\left(
\begin{array}{cccc}
 0 & v^1 & v^2 & v^3 \\
 v^1 & 0 & 0 & 0 \\
 v^2 & 0 & 0 & 0 \\
 v^3 & 0 & 0 & 0 \\
\end{array}
\right) \,, &
&K^1 = -\left(
\begin{array}{cccc}
 0 & v^0 & 0 & 0 \\
 v^0 & 0 & -v^2 & -v^3 \\
 0 & v^2 & 0 & 0 \\
 0 & v^3 & 0 & 0 \\
\end{array}
\right) \,,
\\
&K^2 =-\left(
\begin{array}{cccc}
 0 & 0 & v^0 & 0 \\
 0 & 0 & v^1 & 0 \\
 v^0 & -v^1 & 0 & -v^3 \\
 0 & 0 & v^3 & 0 \\
\end{array}
\right)  \,,& 
&K^3 = - \left(
\begin{array}{cccc}
 0 & 0 & 0 & v^0 \\
 0 & 0 & 0 & v^1 \\
 0 & 0 & 0 & v^2 \\
 v^0 & -v^1 & -v^2 & 0 \\
\end{array}
\right) \,.
\end{aligned}
\end{equation}
These matrices close the algebra~\eqref{Modified_commutator_K_matrices}, where $W^{\mu\nu}$ is given by~\eqref{W-matrix_definition}, with $\tau = 2 \, v^\mu \, v_\mu$ and $\xi =0$. We can see that the W-matrix is zero only in the lightlike cases $v^\mu \, v_\mu = 0$. Notice that in the popular timelike case, $v^\mu = \delta^\mu{}_0$, the matrix $K^0 =0$, while $K^i$ are the three boost generators. This illustrates how $K^\mu$ can be degenerate Lorentz matrices. This is, in fact, even more evident in the 
$\varrho$-Poincar\'e case:
\begin{equation}
\begin{gathered}
f^{\mu\nu}{}_\rho =   \delta^\mu{}_1 \, \delta^\nu{}_0 \, \eta_{\rho 2} -   \delta^\mu{}_2 \, \delta^\nu{}_0 \, \eta_{\rho 1}  \,  \,,
~~
 b^{\mu\nu}{}_\rho = 2 \, \delta^{(\mu}{}_1 \, \delta^{\nu)}{}_0 \, \eta_{\rho 2} -   \delta^{(\mu}{}_2 \, \delta^{\nu)}{}_0 \, \eta_{\rho 1}   \,, 
\\
 c^{\mu\nu}{}_\rho =  2\, \delta^{[\mu}{}_0  \left( \delta^{\nu]}{}_2 \, \delta^1{}_\rho -  \delta^{\nu]}{}_1 \, \delta^2{}_\rho  \right)\,,
\end{gathered}
\end{equation}
where the $K^\mu$ matrices are all zero except $(K^0)^\alpha{}_\beta =  \delta^\alpha{}_1 \, \eta_{\beta 2} -   \delta^\alpha{}_2  \, \eta_{\beta 1}$.

\subsection{Quantum Minkowski spacetimes}\label{Subsec42_MinkowskiSpacetimes}

The quantum Minkowski spacetimes whose isometries are described by one of the quantum groups $C_\ell[ISO(3,1)]$  defined above, will be  $C_\ell[ISO(3,1)]$-comodules (\textit{i.e.} noncommutative representations of  $C_\ell[ISO(3,1)]$), which we will call $C_\ell[\mathbbm{R}^{3,1}]$ as they reduce to the algebra of functions on Minkowski space in the commutative limit. Relations of the RXX type define such comodules:
\begin{equation}\label{RXX_relations_general}
\alpha_1 \,  \x^A \, \x^B + \alpha_2 \, R^{BA}_{CD} \, \x^C \, \x^D = 0  \,,  \qquad \x^A = (\x^\mu , \1)  \in C_\ell[\mathbbm{R}^{3,1}] \,, \alpha_i \in \mathbbm{C} \,.
\end{equation}
Because of the RTT relations~\eqref{RTT_relations}, for any choice of parameters $\alpha_i$, Eqs.~\eqref{RXX_relations_general} are automatically covariant under the following co-action of the quantum group:
\begin{equation}\label{CoactionPoincCoordinates}
    \x'^A = \hat T^A{}_B \otimes \x^B  \qquad  \Rightarrow \qquad  \x'^\mu = \La^\mu{}_\nu \otimes \x^\nu  + \A^\mu \otimes \1  \,, \qquad \1' = \1 \otimes \1\,.
\end{equation}
Replacing the components of the R-matrix previously found, Eq.~\eqref{R_matrix_components_Sol_2}, into Eq.~\eqref{RXX_relations_general} leads to the following constraint, from the $A=4$, $B=\mu$ and $A=\mu$, $B=4$ components:
\begin{equation}\label{Alpha_constraints_1}
    \alpha_1 + \alpha_2 \, (\sigma+\theta)  = 0 \,,
\end{equation}
and, from the $A=\mu$, $B=\nu$ components, to the equation:
\begin{equation}
 [ \x^\mu ,  \x^\nu ] =    i \, \left(   c^{\mu\nu}{}_\rho \,  \x^\rho  +    \theta^{\mu\nu} \,  \1     +  {\sfrac \chi \theta}  \,  \eta^{\mu\nu} \, \1   \right) \,.
\end{equation}
The last term on the right-hand side of the equation is not consistent: when $\mu=\nu$ everything vanishes except for it. We conclude that, in order for a well-define quantum Minkowski space to exist, we need to have
\begin{equation}
    \chi = 0 \,.
\end{equation}
The commutation relations between the spacetime coordinates are then
\begin{equation}\label{XX_commrels}
 [ \x^\mu ,  \x^\nu ] =    i \, \left(   c^{\mu\nu}{}_\rho \,  \x^\rho  +    \theta^{\mu\nu} \,  \1     \right) \,,
\end{equation}
and they can be written in the RXX form as
\begin{equation}
  (\sigma + \theta ) \,   \x^A \, \x^B  =   \,  R^{BA}_{CD}  \,  \x^C \, \x^D \,.
\end{equation}
These commutation relations leads to a consistent associative algebra if they satisfy the Jacobi identities. These turn out to be equivalent to the previously discovered Eqs.~\eqref{JacobiIdentitiesTranslationAlgebra}.

\subsection{Differential calculus}~\label{Subsec43_DiffCalc}

In a similar fashion, one can introduce a 4-dimensional differential calculus as a bicovariant bimodule  $\Gamma_\ell^1$ on $C_\ell[\mathbbm{R}^{3,1}]$~\cite{Woronowicz:1989}, defined by rules of the form
\begin{equation}\label{RXdX_relations}
    \beta_1 \,  \x^A \, d\x^B + \beta_2 \, d \x^A \, \x^B + \, R^{BA}_{CD} \left(   \beta_3 \,  \x^C \, d\x^D + \beta_4 \, d \x^C \, \x^D \right) = 0   \,, 
\end{equation}
where $d \x^A = ( d x^\mu , 0)$ and $\beta_i \in \mathbbm{R}$. These are covariant under the coaction~\eqref{CoactionPoincCoordinates}, extended to the basis one-forms as
\begin{equation}\label{TransformationRuleOneForms}
    d \x'^\mu = \hat T^A{}_B \otimes d \x^B \,, \qquad  \Rightarrow \qquad  d \x^\mu = \La^\mu{}_\nu \otimes d \x^\nu \,,
\end{equation}
Replacing the R-matrix~\eqref{R_matrix_components_Sol_2}, one gets two constraints:
\begin{equation}\label{Beta_constraints}
     \beta_1  +   \theta  \, \beta_3  +  \sigma \, \beta_4 = 0 \,, \qquad
     \beta_2  + \sigma \, \beta_3    + \theta  \, \beta_4  = 0 \,,
\end{equation}
and a set of equations which, once the solution of~\eqref{Beta_constraints} is imposed, reduce to the the following commutation relations:
\begin{equation}\label{BicovariantDifferentialCalculus}
[ \x^{\mu} , d\x^{\nu} ]  = i \,f^{\nu\mu}{}_\rho \, d \x^\rho  \,.
\end{equation}
The final expression of the RXdX relations~\eqref{RXdX_relations}, after solving the constraints on the parameters, is
\begin{equation}\label{RXdX_relations_2}
  -(\theta  \, \beta_3  +  \sigma \, \beta_4)\,  \x^A \, d\x^B + (\sigma \, \beta_3    + \theta  \, \beta_4 ) \, d \x^A \, \x^B + \, R^{BA}_{CD} \left(   \beta_3 \,  \x^C \, d\x^D + \beta_4 \, d \x^C \, \x^D \right) = 0  \,, 
\end{equation}
one gets a two-parameter family of ``RXdX relations'', which all lead to the same commutation relations~\eqref{BicovariantDifferentialCalculus}.

If we impose the Jacobi identities on the products of $d \x^\mu$ with two coordinates:
\begin{equation}
 [\x^{\mu}  ,  [ \x^{\nu} ,  d \x^\rho ]] + [\x^{\nu}  ,  [ \x^{\mu} ,  d \x^\rho ]] + 
[d \x^\rho  , [ \x^\mu ,  \x^\nu ]] 
= \left[  2 \,  f^{\rho[\mu}{}_\sigma \, f^{\sigma|\nu]}{}_\theta +    c^{\mu\nu}{}_\sigma \, f^{\rho\sigma}{}_\theta \right] \, d \x^\theta = 0 \,,
\end{equation}
that is satisfied if
\begin{equation}\label{ModifiedJacobiIdentities_for_f_2}
 f^{\rho\mu}{}_\sigma \, f^{\sigma\nu}{}_\theta   -  f^{\rho\nu}{}_\sigma \, f^{\sigma\mu}{}_\theta   =  - c^{\mu\nu}{}_\sigma \, f^{\rho\sigma}{}_\theta \,,
\end{equation}
which coincides with Eq.~\eqref{ModifiedJacobiIdentities_for_f} if
\begin{equation}
     W^{\mu\nu\rho}_\sigma = \tau \, \delta^{[\mu}{}_\sigma \, \eta^{\nu] \rho} + \xi \, \varepsilon^{\mu\nu\rho}{}_\sigma 
 = 0 \,.
\end{equation}
This is then the condition for the existence of a bicovariant diferential calculus.

The commutation relations~\eqref{BicovariantDifferentialCalculus} imply that the most generic one-form $\omega \in \Gamma_\ell^1$ can be written as
\begin{equation}
\omega = \omega_\mu \, d \x^\mu \,, \qquad \omega_\mu \in C_\ell[\mathbbm{R}^{3,1}] \,.     
\end{equation}
One can define an exterior differential operator, which sends noncommutative scalar functions to one-forms and satisfies the undeformed Leibniz rule
\begin{equation}
    d : C_\ell[\mathbbm{R}^{3,1}] \to \Gamma^1_\ell \,, \qquad d(f \, g) = df \, g + f \, d g \,,
\end{equation}
and a covariant exterior algebra which, again, is undeformed:
\begin{equation}\label{UndeformedExteriorAlgebra}
d \x^\mu \wedge d \x^\nu = - d \x^\nu \wedge d \x^\mu \, \in \Gamma^2_\ell \,, \qquad    d \x^\mu \wedge ( d \x^\nu \wedge d \x^\rho) = (d \x^\mu \wedge d \x^\nu ) \wedge d \x^\rho \,,
\end{equation}
The basis one-forms of $\Gamma^n_\ell$ are the products of $d \x^\mu$, and the most generic $n$-forms can be written
\begin{equation}
\omega_{\mu_1 \dots \mu_n} \, d \x^{\mu_1} \wedge \dots \wedge  d \x^{\mu_n} \in \Gamma^n_\ell \,, \qquad \omega_{\mu_1 \dots \mu_n} \in C_\ell[\mathbbm{R}^{3,1}] \,.
\end{equation}
where $n=1,2,3,4$. The differential will act with the standard graded Leibniz rule on the exterior differential complex $\Gamma^{\wedge}_\ell$:
\begin{equation}
    d(\omega \wedge\rho) = d\omega \wedge \rho + (-1)^m \, \omega \wedge d \rho \,, \qquad \omega \in \Gamma^m_\ell \,, ~~ \rho \in \Gamma^n_\ell \,.
\end{equation}
Notice that the undeformed graded antisymmetry of the wedge product is guaranteed only on basis forms:
\begin{equation}
(d \x^{\mu_1} \wedge \dots \wedge   d \x^{\mu_m})\wedge (d \x^{\mu_1} \wedge \dots \wedge   d \x^{\mu_n})  =   (-1)^{m \, n}
\, (d \x^{\mu_1} \wedge \dots \wedge   d \x^{\mu_n}) \wedge (d \x^{\mu_1} \wedge \dots \wedge   d \x^{\mu_m}) \,,
\end{equation}
while on any other pair of forms, the noncommutativity with the scalar functions  intervenes, \textit{i.e.}
\begin{equation}
 (\omega_\mu \,d\x^\mu ) \wedge d \x^\nu = - \omega_\mu \,d \x^\nu \wedge d\x^\mu  = - d \x^\nu \wedge (\omega_\mu \,d\x^\mu ) - [\omega_\mu , d \x^\nu] \wedge d\x^\mu  \,.
\end{equation}

 The fact that the algebra of the basis forms of the exerior complex $\Gamma^{\wedge}_\ell$ is undeformed is due to the underlying assumption that  the Lorentz matrices are commutative: since the Poincar\'e transformation rule~\eqref{TransformationRuleOneForms} of the basis one-forms does not depend on the translations, a regular graded-commutative wedge product between them will be automatically Poincar\'e-covariant, \textit{e.g.}
 \begin{equation}
     d \x'^\mu \wedge d \x'^\nu = \La^\mu{}_\rho \, \La^\nu{}_\sigma \otimes (d \x^\rho \wedge d \x^\sigma)  = - d \x'^\nu \wedge d \x'^\nu \,.  
 \end{equation}
In general, one has a deformed exterior algebra in which the wedge product between one-forms is controlled by the R-matrix, as follows:
\begin{equation}
d \x^A \wedge d \x^B = - R^{AB}_{CD} \, d \x^C \wedge d \x^D  \,,
\end{equation}
which, when we recall that $d\x^A = \delta^A{}_\mu \, d\x^\mu$, reduces to:
\begin{equation}\label{GeneralRdeformedExteriorAlgebra}
d \x^\mu \wedge d \x^\nu = - R^{\mu\nu}_{\rho\sigma} \, d \x^\nu \wedge d \x^\rho  \,.
\end{equation}
In our case the spacetime components of the R-matrix, $R^{\mu\nu}_{\rho\sigma}$ are undeformed, and the multiplication rules~\eqref{GeneralRdeformedExteriorAlgebra} coincide with~\eqref{UndeformedExteriorAlgebra}. One has a deformed exterior algebra when the Lorentz matrices are noncommutative, which also implies a quadratic noncommutativity between coordinates, see for example~\cite{Podles:1995qy,Meier:2023lku,Meier:2023kzt}.

 Finally, it turns out that, if the invariance of the metric~\eqref{MetricInvariance} holds, the basis maximum-degree form commutes with all coordinates:
  \begin{equation}
  [\x^\mu ,   d \x^0 \wedge d \x^1 \wedge d \x^2 \wedge d \x^3 ] = 0 \,.
 \end{equation}
 This means that the algebra of 4-forms $\Gamma^4_\ell$ and that of 0-forms $C_\ell[\mathbbm{R}^{3,1}]$ are isomorphic, and one can define a Poincar\'e-covariant integration on maximum-degree forms in terms of an integration over noncommutative scalar functions.

\subsection{Braided N-point algebra}\label{Subsec44_Braiding}

Finally, a covariant braided tensor product structure can be introduced,  generalizing the notion of ``multilocal'' functions, and can form the basis for constructing quantum field theories on our quantum Minkowski spacetimes. The most general  ``RXY relations'' are:
\begin{equation}
\label{RXY_relations_general}
   \gamma_1 \,  \x_a^A \,\x_b^B + \gamma_2 \, \x_b^A \, \x_a^B +   R^{BA}_{CD} \left( \gamma_3 \, \x_b^C \, \x_a^D  +  \gamma_4 \, R^{BA}_{CD} \, \x_a^C \, \x_b^D \right) = 0 \,, \qquad \forall a,b \in \mathbbm{N} \,,  
\end{equation}
where the $a$, $b$ indices refer to two (possibly) different points, and $\x_a^A = (\x^\mu_a , \1)$ are all copies of the algebra~\eqref{RXX_relations}.  These rules are covariant under the Poincar\'e coaction on different points:
\begin{equation}
    \x_a'^A = \hat T^A{}_B \otimes  \x_a^B \,,  \qquad  \x_a'^\mu = \La^\mu{}_\nu \otimes  \x_a^\nu + \A^\mu \otimes \1 \,, \qquad  \forall a \in \mathbbm{N} \,.
\end{equation}
Replacing the form~\eqref{R_matrix_components_Sol_2} of the R-matrix, one gets, by inspecting the various components of Eq.~\eqref{RXY_relations_general}, the following two constraints:
\begin{equation}
\gamma_1 +  \theta \,  \gamma_3 +   \sigma  \, \gamma_4  = 0  \,,
\qquad
 \gamma_2 + \sigma \, \gamma_3  + \theta \, \gamma_4 = 0 \,,
\end{equation}
and the equations (recall that we set $\chi = 0$)
\begin{equation}
   \theta   \, [\x_a^{\mu} , \x_b^{\nu}] 
    =  \left( \sfrac{\theta - \sigma }{\theta+\sigma } \right) \left( \sfrac{\gamma_1+\gamma_2}{\gamma_1-\gamma_2} \right) \left[ i \, \theta \, \theta^{\mu\nu}  - i \, {\sfrac \theta 2}  \, c^{\nu\mu}{}_\rho\left(     \x_a^\rho  +    \x_b^\rho     \right) \right] - i \, {\sfrac \theta 2} \, \left( \sfrac{\theta + \sigma }{\theta -\sigma } \right) \left( \sfrac{\gamma_1-\gamma_2}{\gamma_1+\gamma_2} \right) \,   b^{\nu\mu}{}_\gamma \,  \left( \x_a^\rho - \x_b^\rho   \right)   \,.
\end{equation}
We have to demand that, when $a=b$, the algebra above reduces to that of a single point, \textit{i.e.} Eq.~\eqref{XX_commrels}. This happens only if $\gamma_1 = \gamma_0 \, \theta $ and $\gamma_2 = \gamma_0 \, \sigma$ with the same constant $\gamma_0$, and the algebra reduces to
\begin{equation}\label{XY_commrels}
 [\x_a^{\mu} , \x_b^{\nu}] 
    = i \, \theta^{\mu\nu}   
     -  {\sfrac i 2}   \,   b^{\mu\nu}{}_\gamma \,  \left( \x_a^\rho - \x_b^\rho   \right)   
 +   {\sfrac i 2}  \, c^{\mu\nu}{}_\rho\left(     \x_a^\rho  +    \x_b^\rho     \right)    \,.
\end{equation}
and the ``RXY'' form of the algebra is now
\begin{equation}
\label{RXY_relations_2}
   \theta \,  \x_a^A \,\x_b^B +  \sigma \, \x_b^A \, \x_a^B -   R^{BA}_{CD} \,\x_b^C \, \x_a^D    = 0 \,. 
\end{equation}
The algebra~\eqref{XY_commrels} has to satisfy the Jacobi identities:
\begin{equation}
    [\x^\mu_a ,[\x^\nu_b , \x^\rho_c]]  +     [\x^\nu_b , [\x^\rho_c,\x^\mu_a]]
    +     [\x^\rho_c,[\x^\mu_a,\x^\nu_b ]] = 0 \,,
\end{equation}
which, after tedious although elementary calculations, can be put in the form
\begin{equation}
    \begin{aligned}
        &\left( f^{\mu\nu}{}_\sigma \, f^{\sigma\rho}{}_\lambda - f^{\mu\rho}{}_\sigma \, f^{\sigma\nu}{}_\lambda + c^{\rho\nu}{}_\sigma \, f^{\sigma\mu}{}_\lambda \right) \, x^\lambda_a +
        \\
        &\left( f^{\nu\rho}{}_\sigma \, f^{\sigma\mu}{}_\lambda - f^{\nu\mu}{}_\sigma \, f^{\sigma\rho}{}_\lambda + c^{\rho\mu}{}_\sigma \, f^{\nu\sigma}{}_\lambda \right) \, x^\lambda_b +
        \\
        &\left( f^{\rho\mu}{}_\sigma \, f^{\sigma\nu}{}_\lambda - f^{\rho\nu}{}_\sigma \, f^{\sigma\mu}{}_\lambda + c^{\mu\nu}{}_\sigma \, f^{\rho\sigma}{}_\lambda \right) \, x^\lambda_c = 0\,,
    \end{aligned}
\end{equation}
which, of course, coincides with Eq.~\eqref{ModifiedJacobiIdentities_for_f} if
\begin{equation}
     W^{\mu\nu\rho}_\sigma = \tau \, \delta^{[\mu}{}_\sigma \, \eta^{\nu] \rho} + \xi \, \varepsilon^{\mu\nu\rho}{}_\sigma 
      = 0 \,.
\end{equation}

One might notice, in passing, that the difference between two different coordinates, \textit{i.e.},
\begin{equation}
    \Delta \x_{ab}^\mu = \x^\mu_a - \x^\mu_b \,,
\end{equation}
closes the same commutation relations, with a third coordinate $\x^\mu_c$, as the basis one forms do with $\x^\mu$, Eq.~\eqref{RXdX_relations}.  This  is to be expected, as the coordinate differences are invariant under translations, and therefore behave like $d \x^\mu$ under Poincar\'e transformations. Since the commutation relations~\eqref{RXdX_relations} and~\eqref{RXY_relations} are essentially determined by the request of Poincar\'e covariance, the commutators of $\Delta \x_{ab}^\mu$ and $d \x^\mu$ with coordinates turn out to be the same. This also makes sense from the point of view that coordinate differentials are, in some sense, ``infinitesimal'' coordinate differences.

The most important property of the braided algebras I found in the present paper, is that the coordinate differences $\Delta \x_{ab}^\mu$ form a commutative subalgebra of the braided tensor product algebra,
\begin{equation}
    [\Delta \x_{ab}^\mu,\Delta \x_{cd}^\nu] = 0\,, \qquad \forall a,b,c,d \in \mathbbm{N} \,.
\end{equation}
This simplifies enormously the task of making physical sense of a QFT built on one of our quantum Minkowski spacetimes: the theory can be defined in terms of a partition function (which in turn is a noncommutative functional integral), and its physical predictions are encoded in N-point functions calculated as expectation values on products of functions $\phi(\x_{a_1}) \cdot \phi(\x_{a_2})   \dots \phi(\x_{a_n})$ of different points. If these are translationally-invariant (as they should if the QFT is Poincar\'e invariant, barring Poincar\'e breaking/anomalies), then the N-point functions are regular, commutative functions on $\mathbbm{R}^4$, which are invariant under standard/commutative Lorentz transformations.

\subsection{Recap of the Hopf algebras defined thus far}\label{Subsec45_Recap}

The requests I made so far allowed me to identify a sizeable class of Hopf algebras $C_\ell[ISO(3,1)]$ that deform the Poincar\'e group, possessing a quantum Minkowski homogeneous space $C_\ell[\mathbbm{R}^{3,1}]$ and an N-point braiding construction $C_\ell[\mathbbm{R}^{3,1}]^{\otimes_\ell N}$, as well as a 4-dimensional bicovariant differential calculus $\Gamma_\ell^{1}$ and differential complex $\Gamma_\ell^{\wedge}$.

The  assumptions I made were only that the commutation relations were of the form defined by~\eqref{Assumption_1}, \eqref{Assumption_2}  and \eqref{Assumption_3},  and that the homogeneous space, differential calculus and N-point braiding relations satisfied the Jacobi identities (actually, requesting that for the braiding commutators is enough and implies the other two).
The result is a $C_\ell[ISO(3,1)]$ Hopf algebra with undeformed coproduct, counit and antipode~\eqref{Lambda_a_coproducts_counits_antipodes}, and the following algebraic relations:
\begin{equation}
\begin{gathered}
  [ \A^\mu , \A^\nu ]  =  i \, \theta^{\rho\sigma} \,  \left( \delta^{[\mu}{}_\rho \, \delta^{\nu]}{}_\sigma - \La^{[\mu}{}_\rho \, \La^{\nu]}{}_\sigma    \right)  
  + i \, c^{\mu\nu}{}_\rho \, \A^\rho  \,,
\qquad
[ \La^{\mu}{}_\rho ,  \La^{\nu}{}_\sigma ] = 0   \,,
\\
[ \La^{\mu}{}_\rho  , \A^{\nu} ]   =
  i \,  f^{\alpha\beta}{}_\gamma
\left(  \La^{\mu}{}_\alpha \, \La^{\nu}{}_\beta \, \delta^\gamma{}_\rho  - \delta^{\mu}{}_\alpha \, \delta^{\nu}{}_\beta \, \La^\gamma{}_\rho \right) 
 \,,
 \qquad \eta^{\rho\sigma} \,   \La^\mu{}_\rho \, \La^\nu{}_\sigma 
= \eta^{\mu\nu} \,,
~~
\eta_{\mu\nu}  \,   \La^\mu{}_\rho \, \La^\nu{}_\sigma 
= \eta_{\rho\sigma} \,,
 \end{gathered}
\end{equation}
the $C_\ell[\mathbbm{R}^{3,1}]^{\otimes_\ell N}$ algebra of the quantum Minkowski space coordinates (which reduces to the one-point algebra $C_\ell[\mathbbm{R}^{3,1}]$ when $a=b$) is
\begin{equation}
 [\x_a^{\mu} , \x_b^{\nu}] 
    = i \, \theta^{\mu\nu}   
     -  {\sfrac i 2}   \,   b^{\mu\nu}{}_\gamma \,  \left( \x_a^\rho - \x_b^\rho   \right)   
 +   {\sfrac i 2}  \, c^{\mu\nu}{}_\rho\left(     \x_a^\rho  +    \x_b^\rho     \right)    \,.
\end{equation}
and the bicovariant differential calculus $\Gamma_\ell^1$:
\begin{equation}
[ \x^{\mu} , d\x^{\nu} ]  = i \,f^{\nu\mu}{}_\rho \, d \x^\rho  \,.
\end{equation}
The structure constants $ c^{\alpha\beta}{}_\gamma$, $ f^{\alpha\beta}{}_\gamma$ and $\theta^{\alpha\beta}$ satisfy the following constraints
\begin{equation}  \label{ConstraintsRecap}
     \begin{gathered}
      c^{\mu\nu}{}_\lambda \, c^{\lambda\rho}{}_\sigma - 
        \, c^{\mu\rho}{}_\lambda c^{\lambda\nu}{}_\sigma 
        - c^{\nu\rho}{}_\lambda \, c^{\mu\lambda}{}_\sigma  = 0\,,
        \qquad 
 c^{\mu\nu}{}_\lambda \, \theta^{\rho \lambda}   + c^{\rho\mu}{}_\lambda \, \theta^{\nu \lambda} + c^{\nu\rho}{}_\lambda \, \theta^{\mu \lambda} =  0\,,
 \\
 f^{[\alpha\beta]}{}_\gamma = - 
    {\sfrac 1 2 } c^{\alpha\beta}{}_\gamma \,,
    \qquad
    f^{\alpha\mu}{}_\lambda  \, f^{\lambda\nu}{}_\beta -      f^{\alpha\nu}{}_\lambda  \, f^{\lambda\mu}{}_\beta 
    = - c^{\mu\nu}{}_\rho \, f^{\alpha\rho}{}_\beta \,,
    \qquad  
    \eta^{\theta (\sigma}  \,  f^{\varphi ) \lambda }{}_\theta =  0 \,.
     \end{gathered}
\end{equation}
As shown above, these constraints can be solved as follows: consider the four matrices
\begin{equation}
    (K^\mu)^\alpha{}_\beta = f^{\alpha\mu}{}_\beta \,,
\end{equation}
as per Eqs.~\eqref{ConstraintsRecap}, $K^\mu$ are  (possibly degenerate) linear combinations of Lorentz matrices,
\begin{equation}\label{K_as_M}
    K^\mu =  {\sfrac 1 2}   \, \omega^\mu{}_{\rho\sigma} \, M^{\rho\sigma} \,, \qquad  (M^{\mu\nu})^\alpha{}_\beta = \eta^{\mu \alpha} \, \delta^\nu{}_\beta - \eta^{\nu \alpha} \, \delta^\mu{}_\beta \,,
\end{equation}
which close a Lie algebra with structure constants $-c^{\mu\nu}{}_\rho$:
\begin{equation}\label{K_commutation_relations}
    [K^\mu , K^\nu] = - c^{\mu\nu}{}_\rho \, K^\rho \,,
\end{equation}
then, if $c^{\mu\nu}{}_\rho$ satisfy the Jacobi identities, all of the constraints above are satisfied, except the one regarding $\theta^{\mu\nu}$, which needs to be checked case by case: each set of matrices $K^\mu$ might allow for some components of $\theta^{\mu\nu}$ to stay nonzero. 

\section{The Quantum Yang--Baxter Equation} \label{SecQYBE}

The Hopf algebras, homogeneous spaces, differential calculi and braidings I constructed have all the consistency properties that are desirable in order to build classical and quantum field theories on them. However, the question remains whether they are \textit{quasitriangular,} meaning they admit an R-matrix that satisfies the QYBE~\eqref{QYBE}. It is then time to attack the equation, reproduced here for convenience, and check if all the constraints that I found on the form of the R-matrix have hammered it down to a form that allows me to solve it:
\begin{equation}
        R^{AB}_{IJ} \, 
    R^{IC}_{LK} \, 
    R^{JK}_{MN} 
    =
    R^{BC}_{JK} \, 
    R^{AK}_{IN} \, 
    R^{IJ}_{LM} \,,
\end{equation}
the R-matrix found in~\eqref{R_matrix_components_Sol_2}, further constrained by the findings of Sec.~\ref{Sec4_FurtherConstraints}, is given by
\begin{equation}
\begin{aligned}\label{R-matrix_whole_1}
        R^{AB}_{CD} =& \theta \, \delta^{A}{}_C \, \delta^{B}{}_D
 + \sigma \, \delta^{A}{}_D \, \delta^{B}{}_C
-  i  \theta \,  \theta^{\mu\nu} \, \delta^{A}{}_\mu \, \delta^{B}{}_\nu \, \delta^\Q{}_{C} \, \delta^\Q{}_{D}  \\
 & + i \, \sfrac{\theta}{2}  \, \left( b^{\mu\nu}{}_\rho - c^{\mu\nu}{}_\rho \right)  \delta^{A}{}_\mu \, \delta^{B}{}_\nu \, \delta^\rho{}_{C} \, \delta^\Q{}_{D} 
 - i \, \sfrac{\theta}{2}  \, \left( b^{\mu\nu}{}_\rho + c^{\mu\nu}{}_\rho \right) \delta^{A}{}_\mu \, \delta^{B}{}_\nu \, \delta^\Q{}_{C} \, \delta^\rho{}_{D}  \,. 
\end{aligned}
\end{equation}

First, it is convenient to calculate by hand a subset of its components: $A=\Q $, $B=\Q$, $C=\mu$, $L=\Q$, $M=\nu$, $N=\Q$.
\begin{equation}
\begin{gathered}
        R^{\Q \Q }_{IJ} \, 
    R^{I\mu}_{\Q K} \, 
    R^{JK}_{\nu\Q} 
    =
    R^{\Q \mu}_{JK} \, 
    R^{\Q K}_{I\Q } \, 
    R^{IJ}_{\Q \nu} 
    \\
    \Downarrow
        \\
             \bcancel{ (\theta + \sigma)   \,  \theta \, \sigma \, 
\delta^\mu{}_\nu }
    =
   \bcancel{  (\theta + \sigma)  
\,  \theta \, \sigma \, 
\delta^\mu{}_\nu }
    +
     \theta \, \sigma  \, 
\delta^\mu{}_\nu
\end{gathered}
\end{equation}
so we get $ \theta \, \sigma \, 
\delta^\mu{}_\nu =0$. $\theta = 0$ is unacceptable, as it would cancel the identity term in the R-matrix, and the commutative limit would be wrong. We than conclude that $\sigma = 0$. This leaves the parameter $\theta$ as a single normalization factor multiplying all components of $R^{AB}_{CD}$. Since both the RTT relations~\eqref{RTT_relations}  and the QYBE~\eqref{QYBE} are homogeneous in $R^{AB}_{CD}$, this normalization factor is irrelevant and can be set to one, for convenience. Notice that the RXX, RXY and RXdX relations~\eqref{RXX_relations}, \eqref{RXY_relations}, \eqref{RXdX_relations} are all homogeneous in $\theta$ too, after solving the relative constraints on their parameters.
I will then set, from now on,
\begin{equation}
    \theta = 1 \,, \qquad \sigma = 0\,.
\end{equation}

Now, introduce the following two $5\times 5$ matrices:
\begin{equation}\label{PKL_matrices}
 (\mathbbm{P}_\mu)^A{}_B = \delta^{A}{}_\mu \, \delta^\Q{}_{B}\,, \qquad
     (\mathbbm{K}^\mu)^A{}_C =  \sfrac{1}{2}  \, \left(  b^{\nu\mu}{}_\rho - c^{\nu\mu}{}_\rho  \right)  \delta^{A}{}_\nu \, \delta^\rho{}_{C} 
=   \sfrac{1}{2}  \, \left(  b^{\mu\nu}{}_\rho + c^{\mu\nu}{}_\rho  \right)  \delta^{A}{}_\nu \, \delta^\rho{}_{C} 
\,,
\end{equation}
they allow to write the R-matrix~\eqref{R-matrix_whole_1}
as
\begin{equation}\label{R-matrix_whole_2}
 R = \mathbbm{1} \otimes \mathbbm{1} + i \, r \,, \qquad r =   - \theta^{\mu\nu} \,  \mathbbm{P}_\mu \otimes \mathbbm{P}_\nu + \mathbbm{K}^\mu \otimes \mathbbm{P}_\mu   -   \mathbbm{P}_\mu \otimes  \mathbbm{K}^\mu   \,,
\end{equation}
where, naturally, $(\mathbbm{1})^A{}_B = \delta^A{}_B$.
The matrices introduced in~\eqref{PKL_matrices} satisfy the following product rules:
\begin{equation}\label{PKL_products}
           \mathbbm{P}_\mu \,  \mathbbm{P}_\nu = \mathbbm{P}_\mu \, \mathbbm{K}^\nu = 0 \,,
           ~~
   \mathbbm{K}^\alpha \mathbbm{P}_\beta = \sfrac{1}{2}  \, \left(  b^{\lambda\alpha}{}_\beta - c^{\lambda\alpha}{}_\beta  \right)  \, \mathbbm{P}_\lambda 
= \sfrac{1}{2}  \, \left(  b^{\alpha\lambda}{}_\beta + c^{\alpha\lambda}{}_\beta  \right)  \, \mathbbm{P}_\lambda 
=  f^{\lambda \alpha}{}_\beta \, \mathbbm{P}_\lambda \,,
\end{equation}
and this allows one to prove quite easily that $r^2 = 0$. Therefore, the cubic and linear terms in $r$ in the QYBE~\eqref{QYBE} cancel out:
\begin{equation}
r_{12}\, r_{13}\, r_{23} = r_{23}\, r_{13}\, r_{12} = 0 \,,  
\qquad 
r_{12} +  r_{13} + r_{23} = r_{23} + r_{13} + r_{12}  \,, 
\end{equation}
and only the quadratic terms in $r$ survive:
\begin{equation}
\begin{gathered}
        r_{12}\, r_{13}  + r_{12}\, r_{23} + r_{13}\, r_{23} = r_{23}\, r_{13} + r_{23}\,  r_{12} + r_{13}\, r_{12} \,,
        \\
        \Downarrow
        \\
            [ r_{12} , r_{13}]  + [r_{12} , r_{23} ] + [r_{13} , r_{23}] =  0 \,, 
\end{gathered}
\end{equation}
this is the \textit{Classical Yang-Baxter Equation}, which is a cocycle condition on the $r$ matrix.

With a little work, and using the product properties~\eqref{PKL_products}, one can prove that
\begin{equation}
    \begin{aligned}
[ r_{12} , r_{13}]  + [r_{12} , r_{23} ] + [r_{13} , r_{23}] =+ \left( \left[\mathbbm{K}^{\mu },\mathbbm{K}^{\nu }\right] 
+ f^{\nu \mu}{}_\rho \, \mathbbm{K}^{\rho }
- f^{\mu \nu}{}_\rho \, \mathbbm{K}^{\rho } \right) \otimes  \mathbbm{P}_\mu \otimes \mathbbm{P}_\nu  
 \\
+P_\mu \otimes \left(  [\mathbbm{K}^{\nu} , \mathbbm{K}^{\mu }] 
+    f^{\mu \nu}{}_\rho \, \mathbbm{K}^\rho  
- f^{\nu \mu}{}_\rho \, \mathbbm{K}^{\rho } \right)\otimes   \mathbbm{P}_\nu 
\\
+
\mathbbm{P}_\mu \otimes \mathbbm{P}_\nu \otimes \left( [\mathbbm{K}^{\mu }, \mathbbm{K}^{\nu}] 
+ f^{\nu \mu}{}_\rho \,  \mathbbm{K}^{\rho} -     f^{\mu \nu}{}_\rho  \,  \mathbbm{K}^\rho \right)  
\\
+\left( 2 \, \theta^{\sigma\rho } f^{[\mu \nu]}{}_\sigma
 + 2 \, \theta^{\mu \sigma } f^{[\rho \nu]}{}_\sigma 
 - \theta^{\sigma \nu } f^{\mu\rho}{}_\sigma
 -\theta ^{\nu \sigma } f^{\rho \mu}{}_\sigma  
\right) {P}_\mu \otimes {P}_\nu \otimes {P}_\rho\,,
    \end{aligned}
\end{equation}
The linearly independent terms of the equation, that need to vanish, are:
\begin{equation}\label{FinalFormYBE}
    \begin{gathered}
 \theta^{\mu\alpha} \, c^{\beta\gamma}{}_\mu
+ \theta^{\mu\beta} \, c^{\gamma\alpha}{}_\mu
+ \theta^{\mu\gamma} \, c^{\alpha\beta}{}_\mu = 0   \,, \qquad [\mathbbm{K}^\mu  , \mathbbm{K}^\nu ] = - c^{\mu\nu}{}_\rho \, \mathbbm{K}^\rho  \,.
    \end{gathered}
\end{equation}
So far, in this Section, I have not assumed anything regarding the nature of the homogeneous subgroup of the $\La^\mu{}_\nu$ matrices, so the relations above could be applied equally well to the 4-dimensional affine group. Specializing to the Poincar\'e group requires one to impose that $\La^\mu{}_\nu$ are Lorentz matrices, \textit{i.e.}, imposing the covariance of Eq.~\eqref{MetricInvariance}. This has already been done in~\eqref{K_LorentAlgebraMatrices}, and the solution is 
\begin{equation}
    b^{\nu\mu}{}_\rho  - c^{\nu\mu}{}_\rho=       2 \, \eta^{\nu\sigma} \omega^\mu_{\sigma\rho}    \,,
\qquad 
b^{\mu\nu}{}_\rho  + c^{\mu\nu}{}_\rho=       2 \, \eta^{\nu\sigma} \omega^\mu_{\sigma\rho}   \,,
\end{equation}
this implies, for the $\mathbbm{K}^\mu$  matrices:
\begin{equation}
(\mathbbm{K}^\mu)^A{}_B =  \sfrac{1}{2}  \, \left(  b^{\nu\mu}{}_\rho - c^{\nu\mu}{}_\rho  \right)  \delta^{A}{}_\nu \, \delta^\rho{}_B  =   \left( \eta^{\nu\sigma} \omega^\mu_{\sigma\rho}  \right)  \delta^{A}{}_\nu \, \delta^\rho{}_B = \sfrac{1}{2} \, \omega^\mu_{\sigma\rho} \, (\mathbbm{M}^{\sigma\rho})^A{}_B
\,, 
\end{equation}
where
\begin{equation}\label{M_5D_rep}
(\mathbbm{M}^{\mu\nu})^A{}_B = \eta^{\mu A} \, \delta^\nu{}_B - \eta^{\nu A} \, \delta^\mu{}_B \,,
\end{equation}
and therefore the r-matrix takes the following form, parametrized in terms of five antisymmetric matrices, $\theta^{\mu\nu}$  and and $ \omega^\rho_{\mu\nu} $:
\begin{equation}\label{r_matrix_final}
r =   - {\sfrac 1 2} \, \theta^{\mu\nu} \,  \mathbbm{P}_\mu \wedge \mathbbm{P}_\nu + {\sfrac 1 2}   \,   \omega^\mu_{\sigma\rho} \, \mathbbm{M}^{\sigma\rho} \wedge \mathbbm{P}_\mu   \,.
\end{equation}
This is the most generic nilpotent r-matrix on the Poincar\'e algebra. The \textit{triangular} (\textit{i.e.} $[r,r]=0$) r-matrices on the Poincar\'e group are those of interest to us, and they have been classified by Zakrzewski~\cite{Zakrzewski:1994bc,Zakrzewski_1997}. I will reproduce here, in my notation, Zakrzewski's presentation of canonical representatives of each automorphism class, and highlight some cases that are worthy of notice. The r-matrices that concern the present paper are cases 7 to 21 in~\cite{Zakrzewski_1997}.
\begin{itemize}

\item \textbf{Cases 7 \& 8} do not admit a nonzero $\theta$-matrix, in other words the spacetime coordinates close a Lie algebra that does not admit a central extension. Their $r$- matrices are
\begin{equation}
    r_7 =  \eta^{\mu\nu} \, \mathbbm{P}_\mu \wedge \mathbbm{M}_{\nu 0} + \eta^{\mu\nu} \,\mathbbm{P}_\mu \wedge \mathbbm{M}_{\nu 3} + \zeta^{(7)} \, (\mathbbm{P}_0 + \mathbbm{P}_3) \wedge \mathbbm{M}_{12} \,.
\end{equation} 
and
\begin{equation}
    r_8 =  \eta^{\mu\nu} \, \mathbbm{P}_\mu \wedge \mathbbm{M}_{\nu 0} + \eta^{\mu\nu} \,\mathbbm{P}_\mu \wedge \mathbbm{M}_{\nu 3} + \zeta^{(8)}\, (\mathbbm{P}_0 + \mathbbm{P}_3) \wedge (\mathbbm{M}_{01}-\mathbbm{M}_{13}) \,.
\end{equation}
The choice $\zeta^{(7)} =0 $  in case 7 coincides with the lightlike-$\kappa$-Poincar\'e model.

\item \textbf{Cases 9$^{\bm{(s)}}$},\footnote{This case, as reported in Table 1 of Ref.~\cite{Zakrzewski_1997}, is wrong, as the r-matrices for $s\neq 0$ do not satisfy the CYBE. This was corrected in~\cite{tolstoy2007twisted} (pag.~13). I reproduced here the correct r-matrices..} $s= (-1,0,1)$ 
\begin{equation}
\begin{aligned}
    r_{9,s} =~& \theta^{(9)}  \, (\mathbbm{P}_0 + \mathbbm{P}_3) \wedge \mathbbm{P}_2 + [ \mathbbm{P}_1 + s \, (\mathbbm{P}_0 + \mathbbm{P}_3) ] \wedge [\zeta^{(9)}_1( \mathbbm{M}_{01}-\mathbbm{M}_{13} ) +  \zeta^{(9)}_2 (\mathbbm{M}_{23} - \mathbbm{M}_{02}) ]
    + \\
    &\zeta^{(9)}_1 \, (\mathbbm{P}_0 + \mathbbm{P}_3) \wedge  \mathbbm{M}_{03} \,.
\end{aligned}
\end{equation}

\item \textbf{Case 10-16}

These cases have something in common: they all give rise to structure constants $c^{\mu\nu}{}_\rho$ for the spacetime coordinate algebra that define a  \textit{unimodular} Lie group.

\begin{itemize}

\item \textbf{Case 10}
\begin{equation}
    \begin{aligned}
    r_{10} =~& \theta^{(10)}_1 \,  \, (\mathbbm{P}_0 - \mathbbm{P}_3) \wedge \mathbbm{P}_1 + \theta^{(10)}_2 \,  (\mathbbm{P}_0 + \mathbbm{P}_3) \wedge \mathbbm{P}_2 + \mathbbm{P}_1 \wedge (\mathbbm{M}_{23}-\mathbbm{M}_{02}) +\\
    &(\mathbbm{P}_0 + \mathbbm{P}_3) \wedge (\mathbbm{M}_{01}-\mathbbm{M}_{13}) \,.
    \end{aligned}
\end{equation}

\item \textbf{Case 11}
\begin{equation}
    r_{11} =\theta^{(11)}_1 \,  \, (\mathbbm{P}_0+\mathbbm{P}_3) \wedge \mathbbm{P}_1 + \theta^{(11)}_2 \, (\mathbbm{P}_0-\mathbbm{P}_3)\wedge \mathbbm{P}_2 + \mathbbm{P}_2 \wedge ( \mathbbm{M}_{01} - \mathbbm{M}_{13} )   \,,
\end{equation}

\item \textbf{Case 12}\footnote{Ref.~\cite{Zakrzewski_1997} has a second mistake: as pointed out in~\cite{tolstoy2007twisted} (pag. 15), the r-matrix of Case 12 in Table 1 has an additional term $(\mathbbm{P}_0-\mathbbm{P}_3) \wedge \mathbbm{P}_2$, which does not satisfy the CYBE. This term has been removed here.}
\begin{equation}
\begin{aligned}
    r_{12} =~&   (\mathbbm{P}_0-\mathbbm{P}_3) \wedge \left[ \theta^{(12)}_1 \, (\mathbbm{P}_0+\mathbbm{P}_3) + \theta^{(12)}_2 \, \mathbbm{P}_1  \right]  + 
    \\
    &(\mathbbm{P}_0+\mathbbm{P}_3) \wedge  \left[\theta^{(12)}_3 \,    \mathbbm{P}_2  +  ( \mathbbm{M}_{01} - \mathbbm{M}_{13} )   \right] \,,
\end{aligned}
\end{equation}

\item \textbf{Case 13, 14 \& 15}

These three cases are the $\varrho$-Poincar\'e-like models found by Lukierski and Woronowicz in~\cite{Lukierski:2005fc}. They divide into three automorphism classes, according to whether the coordinate that acts on $\x^1$ and $\x^2$ as a rotation is timelike (here chosen to be $\x^0$, as a representative case of the class - notice that this is the model that was studied in~\cite{Fabiano:2023uhg}), spacelike ($\x^3$) or lightlike ($\x^0+\x^3$). The family of r-matrices is 
\begin{equation}
    r_{\varrho} =  \theta^{(\varrho)}_1 \, \mathbbm{P}_0 \wedge \mathbbm{P}_3 +  \theta^{(\varrho)}_2 \, \mathbbm{P}_1 \wedge \mathbbm{P}_2  +  \zeta^{(\varrho)}_\mu \, \mathbbm{P}^\mu \wedge  \mathbbm{M}_{12}    \,,
\end{equation}
where $\zeta^{(\varrho)}_\mu $ can be $\delta^0{}_\mu$, $\delta^0{}_\mu+\delta^3{}_\mu$ or $\delta^3{}_\mu$, in the three cases.
Notice that the r-matrices above include a two-parameter family of central extensions of the original models presented by Lukierski and Woronowicz, parametrized by $\theta^{(\varrho)}_1$ and $\theta^{(\varrho)}_2$.

\end{itemize}

\item \textbf{Case 16}
\begin{equation}
    r_{16} =  \theta^{(16)}_1 \, \mathbbm{P}_0 \wedge \mathbbm{P}_3 +  \theta^{(16)}_2 \, \mathbbm{P}_1 \wedge \mathbbm{P}_2  +  \mathbbm{P}_1 \wedge  \mathbbm{M}_{03}    \,.
\end{equation}

\item \textbf{Case 17}
\begin{equation}
    r_{17} =  \theta^{(17)}_1 \, \mathbbm{P}_1 \wedge \mathbbm{P}_2 +  \theta^{(17)}_2 \, (\mathbbm{P}_0 + \mathbbm{P}_3) \wedge \mathbbm{P}_1  +  (\mathbbm{P}_0 + \mathbbm{P}_3) \wedge  \mathbbm{M}_{03}    \,.
\end{equation}

\item \textbf{Case 18}
\begin{equation}
    r_{18} =  \theta^{(18)} \, \mathbbm{P}_1 \wedge \mathbbm{P}_2 
     +  (\mathbbm{P}_0 + \mathbbm{P}_3) \wedge \left(  \mathbbm{M}_{03}  + \zeta^{(18)} \,  \mathbbm{M}_{12} \right)   \,.
\end{equation}

\item \textbf{Cases 19, 20 \& 21}

These cases are simply $\theta$-Poincar\'e, which Zakrzewski calls ``soft deformations''. In~\cite{Zakrzewski_1997} these are divided into three automorphism classes:
\begin{equation}
r_{19} = \mathbbm{P}_1 \wedge (\mathbbm{P}_0 + \mathbbm{P}_3) \,,
\qquad
r_{20} = \mathbbm{P}_1 \wedge \mathbbm{P}_2 \,,
\qquad
r_{21} = \mathbbm{P}_0 \wedge  \mathbbm{P}_3 + \theta^{(21)} \, \mathbbm{P}_1 \wedge \mathbbm{P}_2 \,.
\end{equation}
\end{itemize}

Notice that all cases above that exhibit terms of the type $ \mathbbm{P}_\mu \wedge \mathbbm{P}_\nu $ in the r-matrix do not depend on $\theta^{\mu\nu} $ being nonzero. In other words, these terms are always multiplied by some parameters that we are free to send to zero. This is because the $\theta^{\mu\nu}$ matrix in~\eqref{r_matrix_final} is a solution of the second of Eqs.~\eqref{JacobiIdentitiesTranslationAlgebra}, which depends on the structure constants $c^{\mu\nu}{}_\rho$. These structure constants need to satisfy two constraints that do not involve the $\theta^{\mu\nu}$ matrix at all: 
the Jacobi identities and Eq.~\eqref{ModifiedJacobiIdentities_for_f}. For each solution of these two equations, one should check which components of the $\theta^{\mu\nu}$ matrix can be nonzero in accordance with Eq.~\eqref{JacobiIdentitiesTranslationAlgebra}, which states that $\theta^{\mu\nu}$ is a 2-cocycle for the Lie algebra with structure constants $c^{\mu\nu}{}_\rho$. So we ended up with a classification of 16 quantum groups of isometry of noncommutative spacetimes with commutators that close a Lie algebra, together with all the possible central extensions of these Lie algebras.

\section{Conclusions}

In this paper, inspired by the specific form of the commutators of the $\theta$-, (lightlike) $\kappa$- and $\varrho$-Poincar\'e quantum groups, I set out to find what is the most general quantum Poincar\'e group that possesses all the good properties of these three cases. 
The first of these properties is that the Lorentz subgroup remains commutative, which is something that can be justified on dimensional grounds. The second is that the translation sector is specified by an integrable system of vector fields on the Lorentz group, like in Majid's bicrossproduct construction~\cite{Majid:1994cy}, although I did not assume that the translations close a Lie algebra among each other, allowing for a more general construction. Moreover, I required to be able to define the quantum homogeneous spacetime and a 4-dimensional differential calculus as, respectively, a comodule over the quantum Poincar\'e group and a bicovariant bimodule over the spacetime coordinate algebra. This ensures the existence of a covariant braided tensor product representation of the coordinate algebra, which possesses all the desirable properties for the algebra of N points, in particular the commutativity of coordinate differences, which imply that the N-point functions of any QFT built on the noncommuative spacetime will be ordinary, commutative distributions. 

Imposing all my requirements on the most general product expressed in terms of RTT relations~\eqref{RTT_relations}, I am able to pin down the most general quantum Poincar\'e group in terms of 70 parameters, organized into ``generalized structure constants'' $f^{\mu\nu}{}_\rho$, and an antisymmetric matrix $\theta^{\mu\nu}$. The antisymmetric part of $f^{\mu\nu}{}_\rho$ is proportional to the structure constants $c^{\mu\nu}{}_\rho$ of the Lie-algebra-like  part of the coordinate commutators, while the symmetric part appears  in the commutators between translations and Lorentz matrices, in the differential calculus and in the braided N-point algebra. The generalized structure constants have to satisfy three constraints equations~\eqref{ConstraintsRecap}, which force them to be  (not necessarily linearly independent) linear combinations of Lorentz algebra matrices (in the 4-dimensional representation), which need to close an algebra~\eqref{K_commutation_relations} with structure constants $-c^{\mu\nu}{}_\rho$. The matrix $\theta^{\mu\nu}$ has to satisfy the constraints for a central exension of the Lie algebra with structure constants $c^{\mu\nu}{}_\rho$~\eqref{ConstraintsRecap}, \textit{i.e.}, it needs to be a 2-cocycle. Any solution of these constraints, it turns out, corresponds to a triangular R-matrix, solution of the QYBE, which describes the quantum Poincar\'e group through RTT relations~\eqref{RTT_relations}, the quantum homogeneous spacetime through RXX relations~\eqref{RXX_relations}, the differential calculus through Eqs.~\eqref{RXdX_relations_2}, and the braiding through RXY relations~\eqref{RXY_relations}.
As it turns out, the most general R-matrix satisfying all of the constraints is completely fixed by a triangular r-matrix, which, being nilpotent, is linearly related to the R-matrix. This r-matrix is the most general triangular r-matrix on the Poincar\'e group that does not involve terms of the form $M_{\mu\nu} \wedge M_{\rho\sigma}$, and these have been completely classified by Zakrzewski in~\cite{Zakrzewski:1994bc,Zakrzewski_1997}. The end result is a set of 16 models which are inequivalent under automorphisms, although they can be reorganized into at most 10 categories that are qualitatively different. The picture that emerges is that of a rich landscape of models that is ripe for exploration. For clarity, I decided to call this class of models T-Minkowski spacetimes, and their quantum isometry groups T-Poincar\'e, where the T refers to the translation sector, which is where all of the nontriviality of the models is concentrated.

The upcoming second paper in this series~\cite{Mercati:2024rzg} will focus on the general features of field theory on the T-Minkowski noncommutative spacetimes, in a model-independent fashion.

\section*{Acknowledgements}

I acknowledge  support by the Agencia Estatal de Investigaci\'on (Spain) under grants CNS2023-143760, PID2019-106802GB-I00/AEI/10.13039/501100011033,
and by the Q-CAYLE Project funded by the Regional Government of Castilla y Le\'on (Junta de Castilla y Le\'on) and by the Ministry of Science and Innovation MICIN through NextGenerationEU (PRTR C17.I1).

\end{document}